\documentclass[journal,11pt,graphicx]{IEEEtran}

\newcommand{\lb}{[\![}
\newcommand{\rb}{]\!]}

\newcommand{\frametext}[1]
{
\noindent
{\small
\parbox[t]{7.6cm}
{#1 } }

\vspace{0.1cm} }

\newcommand{\omx}[1]{\overline{\mbox{\sc #1}}}
\newcommand{\wmx}[1]{\widetilde{\mbox{\sc #1}}}

\usepackage{amsmath}
\usepackage{amssymb}
\usepackage{fullpage}
\usepackage[all]{xy}

\newcommand{\remove}[1]{}

\newtheorem{definition}{Definition}

\newtheorem{lemma}{Lemma}
\newtheorem{theorem}{Theorem}

\newtheorem{corollary}{Corollary}

\begin{document}
 
\title{Key-and-Signature  Compact Multi-Signatures for Blockchain: A Compiler with  Realizations}

\author{Shaoquan Jiang, Dima Alhadidi, Hamid Fazli Khojir
\thanks{All authors are   with School of Computer Science,
University of Windsor,
 Windsor, ON N9B 3P4.
Email: jiangshq@uwindsor.ca} }
\maketitle

\begin{abstract}
Multi-signature is a protocol where a set of signatures jointly sign a message so that the final signature is significantly shorter than concatenating individual  signatures together. Recently, it finds applications in blockchain, where several users want to jointly authorize a payment through a multi-signature.  However, in this setting, there is no centralized authority and it could suffer from a rogue key attack where the attacker can generate his own keys arbitrarily.
Further, to minimize the storage on blockchain, it is desired that the aggregated  public-key and the aggregated signature are both as short as possible. In this paper,  we find a compiler that converts a kind  of identification (ID) scheme (which we call a linear ID) to a multi-signature so that both the aggregated  public-key and the aggregated signature have a  size independent of the number of signers. Our compiler is provably secure. The
advantage of our results is that we reduce a multi-party problem to a weakly secure two-party problem.  We realize our compiler with two ID schemes. The first is  Schnorr ID. The second is a new lattice-based ID scheme,  which via our compiler gives the first regular lattice-based multi-signature scheme with  key-and-signature compact without a restart during signing process.
\end{abstract}
\section{Introduction}

A multi-signature scheme allows a group of signers to jointly generate a signature while  no subset of them  can represent all the members to  generate it.  It was first introduced by Itakura and
Nakamura \cite{IN83}.  A trivial method is to ask each signer to generate a signature on the message and concatenate their signatures together. However, this is not efficient: (1) the signature size is linear in the number of signers $n$; (2) we need to provide $n$ signer public-keys to verifier; (3) the verification needs to verify $n$ signatures; (4) all the $n$ public-keys need to be provided to the verifier; (5) the communication and storage complexity for the signature  are both linear in $n$. With applications in blockchain, these problems are crucial   as  the signature will be transmitted, verified and stored in the  blockchain network.  Hence, it is desired to find multi-signature that has a signature with these efficiency measure  independent of $n$.

Early multi-signarture schemes \cite{Harn94,LHL95,OO91} assumed the signer keys are chosen honestly. In Bitcoin \cite{Nak08}, every user can choose his own public-key. However, this might raise a very serious issue. For example, if a user wants to generate a multi-signature with users of  3 pubic-keys $g^{x_1}, g^{x_2}, g^{x_3}$, he could choose $s$ randomly and compute his public-key as $pk=g^{s}(g^{x_1+x_2+x_3})^{-1}$. If the aggregated  public-key (which is the only public-key provided to the verifier) is the multiplication of the four public-keys, then attacker knows its secret and hence can forge a multi-signature.  This is called {\em a rogue key attack.}  How to construct a key-and-signature compact multi-signature scheme secure against a rogue key attack is an important question.

\subsection{Related Works} A multi-signature scheme \cite{IN83} is  a special case of aggregate signature \cite{BGLS03} where each signer of the latter can  sign a possibly different message.  In this work, we only discuss a multi-signature scheme with a motivation of blockchain application where the public-key is arbitrary and the target is to minimize the signature and the aggregated  public-key size. Micali et al. \cite{MOR01} requires an interactive key generation among signers and hence is not suitable.  Boldyreva \cite{Bold03} and Lu et al. \cite{LOS+06} require signers to add {\em proof of possession} (PoP)  to their public-keys, which is typically a signature of the user's  public-key.   The main disadvantage of this assumption is the increase of the public-key size. In the signing process, it also requires a signer to verify the PoP of all the other signers. In addition, this assumption  is not compatible with an ordinary signature where PoP is not required.

Bellare and Neven \cite{BN06} converted the Schnorr signature \cite{Schnorr} into a multi-signature by linearly adding the signature together. Their protocol is of 3-round but without the aggregated key aggregation. Bagherzandi et al. \cite{BCJ08}, Ma et al. \cite{Ma+10}, Syta et al. \cite{ST+16}  and Maxwell et al. \cite{MP+18} attempted to construct a 2-round multi-signature scheme which essentially tries to remove the preliminary committing message which is a hash of the first message in an ID scheme (see \cite{BN06} for example). However, Drijvers et al. \cite{DE+19} pointed out that all these schemes have proof flaws.  They then proved that a slightly modified scheme of Bagherzandi et al. \cite{BCJ08} is secure under the PoP assumption. Other 2-round proposals that support  the key-and-signature  aggregation are due to Alper and Burdges \cite{AB21} and  Nick et al. \cite{NRS21, NRSW20}, where Nick et al. \cite{NRSW20} employed a generic NIZK proof while the other two proposals \cite{AB21,NRS21} are efficient in aggregated key and signature and verification cost (similar to the original Schnorr signature). Boneh et al. \cite{BDN18} proved the security of a modified version of  Maxwell et al. \cite{MP+18} via an added preliminary committing message and hence it is a 3-round scheme.  Bellare and Dai \cite{BD21} proposed a 2-round multi-signature scheme with a tight reduction without the key aggregation.

\begin{figure}
\label{fig:comp} {\small
\begin{center}
\begin{tabular}{|c|c|c|c|c|c|}
\hline
             & Key    & Round         &  $\sharp$ & Assump/\\
       & Compact          & Comp      & Restart  & Remark \\
\hline
\cite{FH19} & {No} & 3 & $\textsf{exp}$ & R-LWE \\
\hline
\cite{FH20}  & {No} & 3  & $\textsf{exp}$ & non-standard\\
\hline
\cite{BTT22} & {\bf Yes} & 2 & $\textsf{exp}$ & R-MLWE\& \\
&&&& R-MSIS\\
\hline
\cite{DOTT21}  & {No} & 2 & $\textsf{exp}$ & R-MLWE\& \\
&&&& R-MSIS\\
\hline
\cite{FSZ22} & {No} & 1 & {\bf 0} & R-SIS\\
&&&& limited-sign \\
\hline
ours & {\bf Yes} & 3 & {\bf 0} & R-SIS \&\\
&&&& R-LWE \\
\hline
\end{tabular}
\end{center}
}
\caption{\small Performance of Lattice-based Secure Multi-Signature Schemes: compact means the size independent of $\sharp$ signers; all schemes have compact signatures; schemes requiring a honest key generations are not included; $\sharp$ restart is  $\sharp$ of repeated runs of signing algorithm (in case it aborts); \textsf{exp} means exponential in either $\sharp$ signers or the security parameter; limited-sign restricts each user to have  a bounded  number of signings. }
\label{fig: compare}
\end{figure}
The above constructions are all based on variants of  the discrete logarithm assumption.
It is important to find out quantum-resistant schemes while this is not easy. For instance, lattice-based scheme \cite{KD20} is insecure \cite{LTT20}. Also,   the proof for a ring-SIS  based scheme  \cite{KSD21} is invalid.  They reduced to find a short $W$ for ring-SIS problem $AW=0$ with public parameter $A$. However,  their obtained $W$ is  trivially zero which does not contradict the ring-SIS assumption. Some schemes \cite{FH19,FH20,MJ19,BTT22} need an exponential number of restarts of the signing algorithm,  due to a noticeable probability of abort event. Some schemes \cite{EB16,MJ19} are provably secure only when all the user keys are   generated honestly which is not suitable for blockchain.
Damg{\aa}rd et al. \cite{DOTT21} and Fleischhacker et al. \cite{FSZ22} do not support key aggregations while  the latter can only allow a signer to sign a predefined number of signatures.
Thus, currently no  multi-signature scheme can support  a key-and-signature aggregation without a restart and allow an unlimited number of signing. Our work is to study this question in details.

\subsection{Contribution}
In this paper, we consider the key-and-signature compact  multi-signature. That is, both key and signature support aggregation and  have a size independent of the number of signers. Toward this, we formulate the {\em linear}   identification scheme (ID)  and propose a compiler that transforms a linear ID to a key-and-signature compact multi-signature scheme, where the signature size and the aggregated public-key are independent of the number of signers.  The advantage of our compiler is that we reduce the multi-party signature problem to a two-party identification problem and hence it is much easier to deal with  and also the security proof for latter  should be simpler. We formulate the linearity of ID via the ${\cal R}$-module from algebra. Our compiler  is provably secure.  We realize our compiler with two ID schemes.  The first is Schnorr ID scheme. The second one is a new ID scheme over ring that is secure under ring-LWE and ring-SIS assumptions. Our ID scheme via the compiler gives the first key-and-signature compact multi-signature without a restart  during the signing process (see Fig. \ref{fig: compare} for a comparison with other schemes), where a signer can do an unlimited number of signing (unlike \cite{FSZ22}, which can only  do a predetermined number of signings).  The security of ID schemes is formulated in terms of unforgeability against an aggregated key of  multi-users with at least one of them honest. Our ID schemes are proven secure through a new forking lemma (called nested forking lemma). Our forking algorithm has a nested rewinding and is more effective  than the previous algorithms which fork at two or more spots sequentially.

\section{Preliminaries} \label{sect: Pre}

\noindent {\bf Notations. } We will use the following notations.
\begin{itemize}
\item  $x\leftarrow S$ samples $x$ uniformly random from a set $S$.
\item For a randomized  algorithm $A$, $u= A(x; r)$ denotes the output of $A$ with input $x$ and randomness $r$,  while $u
\leftarrow A(x)$ denotes the random output (with unspecified randomness).
\item We use $P_R(r)$ to denote the probability $\Pr(R=r)$; for Boolean variable $G$, $\Pr(G)$ means $\Pr(G=1)$.
\item  PPT stands for probabilistic polynomial time.
\item  Min-entropy $H_\infty(X)=-\log (\max_x\log P_X(x))$.
\item $A|B$ stands for $A$ concatenating with $B$.
\item \textsf{negl}$(\lambda)$ is {\em negligible}:   $\lim_{\lambda\rightarrow \infty} poly(\lambda)\textsf{negl}(\lambda)=0$ for any polynomial $poly(\lambda).$
\item     $[\nu]$ denotes set $\{1, \cdots, \nu\}.$
\end{itemize}

\subsection{Ring and  Module} \label{sect: R-module}
In this section, we review a math concept:  module  (for details, see \cite{Lang02}). We start with the concept of ring.    A {\bf ring} $A$  is a set, associated  with multiplication and addition operators,  respectively written as a product and a sum, satisfying the following conditions:
\begin{itemize}
\item[-] {\bf R-1. }  $A$ is a commutative group under addition operator $+$ with identity element {\bf 0}.
\item[-] {\bf R-2. } $A$ is associative under multiplication operator:  for $a, b, c\in A$, (ab)c=a(bc). Also, it has a unit element {\bf 1}: {\bf 1}a=a.

\item[-] {\bf R-3. } It satisfies the  distributive law: for $a, b, c\in A$, $a(b+c)=ab+ac$ and $(b+c)a=ba+ca.$
\end{itemize}
In this paper,  we only consider a {\em commutative ring}:  if $a, b\in A$, then $ab=ba$. That is, when we say ring, it always means a commutative ring.  Note  that a non-zero element in a ring  does not necessarily have a (multiplicative)  inverse, where $b$ is an inverse of $a$ if $ab={\bf 1}$. For instance, in $\mathbb{Z}_{10}$, 3 is an inverse of 7 while $5$ does not have an inverse. If $A$ is a commutative ring with ${\bf 0}\ne {\bf 1}$ and every  non-zero element in $A$ has an inverse, then  $A$ is a {\bf field}.

Now  we introduce the concept {\em module}.

\vspace{0.05in}
\begin{definition} Let $R$ be a ring. An Abelian  group  $M$ (with group operator $\boxplus$) is a {\bf $R$-module}, if (1)  it has defined a multiplication operator $\bullet$ between $R$ and $M$: for any $r\in R, m\in M$, $r\bullet m\in M$; (2) the following conditions are satisfied: for any $r, s\in R$ and $x, y\in M$,
\begin{itemize}
\item[1.] $r\bullet (x\boxplus y)=(r\bullet x)\boxplus (r\bullet y)$;

\item[2.] $(r+ s)\bullet x=(r\bullet x)\boxplus (s\bullet x)$

\item[3.] $(rs)\bullet x=r\bullet (s\bullet x)$

\item[4.]  ${1}_R\bullet x=x$, where ${1}_R$  is  the multiplicative identity of $R$.
\end{itemize}
\end{definition}

\vspace{.10in} We remark that the group operator $\boxplus$ for $M$ is not necessarily the regular  number   addition (e.g., it can be the integer multiplication).

In the following, we give some $R$-module examples.

\vspace{.10in} \noindent {\bf Example 1. } Let $q$ be  a prime and $M$ is a group of order $q$ with generator $g$ (i.e., $M=\langle g\rangle$). Examples of $M$ are a subgroup of $\mathbb{Z}_p^*$ or an elliptic curve group. $x, y\in M$, $xy$ denotes its group operation.  Then, $M$ is a $\mathbb{Z}_q$-module  with  $\bullet$ defined as  $r\bullet m\stackrel{def}{=} m^r$,  for $r\in \mathbb{Z}_q$ and $m\in M$.  It is well-defined: since $m^q=1$,   any representative $r$ in $\mathbb{Z}_q$ such as $r, r+q$  gives the same result $r\bullet m$.   For $r, s\in \mathbb{Z}_q$ and $x, y\in M$,  we check  the module conditions: (1) $s\bullet (xy)=(xy)^s=x^s y^s=(s\bullet x)(s\bullet y); $ (2) $(r+s)\bullet m=m^{r+s}=m^r m^s=(r\bullet m)(s\bullet m)$; (3) $(rs)\bullet x=x^{rs}=(x^s)^r=r\bullet (s\bullet x)$; (4) ${\bf 1}\bullet x=x^1=x.$

\vspace{.10in} \noindent {\bf Example 2. } For any integer $n>0$, $M=\mathbb{Z}_n$ (as an additive group) is a $\mathbb{Z}_n$-module, where $\bullet$ is simply the modular multiplication. The verification of module properties is straightforward.

\vspace{.10in} \noindent {\bf Example 3. } Let $n$ be a positive integer. Then,  the polynomial ring $M=\mathbb{Z}_n[x]$ (as an additive group) is a $\mathbb{Z}_n$-module with $\bullet$ being the modular $n$ multiplication: for $s\in \mathbb{Z}_n, m=\sum_{i=0}^t u_i x^i$, $s\bullet m=\sum_{i=0}^t u_i s x^i$, where $u_is$ is the multiplication over $\mathbb{Z}_n$.
All the other verifications of the properties are straightforward.

\section{Nested Forking Lemma}
The original forking lemma was formulated by Pointcheval and Stern \cite{PS00} to analyze Schnorr signature \cite{Schnorr}. It basically shows that if the attacker can forge a Schnorr signature in the random oracle model \cite{BR93} with a non-negligible probability, then it can generate two forgeries when reminding to the place where the random oracle value was revised. Bellare and Neven \cite{BN06} generalized the forking lemma to a general algorithm \textsf{A}, without resorting to a signature scheme.  This was further generalized by Bagherzandi et al.  \cite{BCJ08} so that \textsf{A} is rewound to many places. However, the algorithm needs $O(n^2 q/\epsilon)$ rewindings, where $q$ is the number  of random values in one run  of $\textsf{A}$ (which is the number  of random oracle queries in typical cryptographic applications)  and $\epsilon$ is the successful probability of $\textsf{A}$ while $n$ is the number of rewinding spots. However, this is not efficient and can even be essentially exponential for a non-negligible $\epsilon$.  The main issue comes from the fact the  rewinding for each spot is  repeated independently until a new success is achieved. But it does not relate different  rewindings. In this section, we give a new forking lemma for two rewinding spots (say at index $i, j$ with $i <j$) while it can be generalized to $n$ rewinding spots.  The new feature here is that
the rewinding is {\em nested}. To see this, suppose that  the first run of $\textsf{A}$ uses the list of random values: $h_1, \cdots, h_{i-1}, h_i, \cdots, h_{j-1}, h_j,\cdots, h_q$ and the rewinding spots are chosen at  index $i$ and $j$. Then,  we execute  \textsf{A} for  another 3 runs with  rewindings that respectively  the following lists of random values:
\begin{align}
&h_1, \cdots, h_{i-1}, h_i, \cdots, h_{j-1}, h_j',\cdots, h_q';  \label{eq: fork0-1} \\
&h_1, \cdots, h_{i-1}, \bar{h}_i, \cdots, \bar{h}_{j-1}, \bar{h}_j,\cdots, \bar{h}_q; \label{eq: fork0-2} \\
&h_1, \cdots, h_{i-1}, \bar{h}_i, \cdots, \bar{h}_{j-1}, \underline{h}_j,\cdots, \underline{h}_q'. \label{eq: fork0-3}
\end{align}
That is, execution (\ref{eq: fork0-1})  rewinds the initial execution to index $j$; execution (\ref{eq: fork0-2}) rewinds the initial execution to index $i$ while   execution (\ref{eq: fork0-3}) rewinds the (rewound) execution (\ref{eq: fork0-2}) to index $j.$ With these related executions, we are able to claim  the outputs are all successful with probability at least $O(\epsilon^4),$  which is still non-negligible.  The advantage of this nested  forking  is that it can be {\em directly} used to extract a secret hidden in recursive random oracle evaluations. Our algorithm will use  the following notations.

 $h\lb 1, \cdots, q\rb\stackrel{def}{=}h_1, \cdots, h_q$ (a sequence of elements);

 $h\lb 1, \cdots, \widehat{i, \cdots, q}\rb\stackrel{def}{=}h_1, \cdots, h_{i-1}, \hat{h}_i, \cdots, \hat{h}_q; $

$h\lb 1, \cdots, \widehat{{i, \cdots, j}}, \overline{{j+1, \cdots, q}}\rb$\\
$\mbox{\hspace{.05in}} $   ${=}h_1 \cdots, h_{i-1}, \hat{h}_i, \cdots, \hat{h}_j, {\bar{h}}_{j+1}, \cdots, {\bar{h}}_q.$

\noindent Other variants such as
 $h\lb 1, \cdots, \overline{i, \cdots, j}, \underline{j+1, \cdots, q}\rb)$
can be defined similarly. Our forking algorithm is in Fig. \ref{fig: fork}.
\begin{figure}[!ht]
\frametext{
----------------------------------------------------------------\\
{\bf Algorithm} $F_{\textsf{A}}(x)$\\
----------------------------------------------------------------

 pick coin $\rho$ for $\textsf{A}$ at random

 $h_1, \cdots, h_q\leftarrow H$

 $(I_0, J_0, \sigma_0)\leftarrow \textsf{A}(x, \ h\lb 1, \cdots, q\rb; \ \rho)$

 If $I_0=0$ or $J_0=0$ or $I_0\ge  J_0$, return \textsf{Fail}

 $\hat{h}_{J_0}, \cdots, \hat{h}_q\leftarrow H$

$(I_{1}, J_{1}, \sigma_{1})\leftarrow \textsf{A}(x,\  h\lb 1, \cdots, \widehat{J_0, \cdots,  q}\rb;\  {\rho})$

 If $I_{1}=0$ or $J_{1}=0$, return \textsf{Fail}

 $\bar{h}_{I_0}, \cdots, \bar{h}_q\leftarrow H$

 $({I}_{2}, {J}_{2}, {\sigma}_{2})\leftarrow \textsf{A}(x,\  h\lb 1, \cdots, \overline{I_0, \cdots, q} \rb; \  \rho)$

 If ${I}_{2}=0$ or ${J}_{2}=0$, return \textsf{Fail}

 $\underline{h}_{J_0}, \cdots, \underline{h}_q\leftarrow H$

\noindent $({I}_3, {J}_3, {\sigma}_3)\leftarrow \textsf{A}(x, h\lb{1, \cdots}, \overline{I_0, \cdots, J_0-1}, \underline{J_0, \cdots,  q}\rb;\rho)$

 If ${I}_3=0$ or ${J}_3=0$, return \textsf{Fail}

 Let {\small $\textsf{Flag}_1=(I_0=I_1=I_2=I_3) \wedge (J_0=J_1=J_2=J_3)$}

 Let $\textsf{Flag}_2=(h_{I_0}\ne \bar{h}_{I_0}) \wedge (h_{J_0}\ne \hat{h}_{J_0})\wedge (\bar{h}_{J_0}\ne \underline{h}_{J_0})$

If $\textsf{Flag}_1\wedge\textsf{Flag}_2$, return $(I_0, J_0, \{\sigma_i\}_{i=0}^3)$

 else return \textsf{Fail}. \\
----------------------------------------------------------------
}
\caption{Forking Algorithm $F_{\textsf{A}}$} \label{fig: fork}
\end{figure}

Before introducing our lemma,  we give two facts.

\vspace{.10in} \noindent {\bf Fact 1. }  {\em For any random variable $I, R$ and any function $F()$ on $I, R$, we have}
\begin{equation*}
\Pr(I=i\wedge F(I, R)=f)=\Pr(I=i\wedge F(i, R)=f).
\end{equation*}
\noindent {\bf Proof. }  For any function $G$ and any random variable $W$, $\Pr(G(W)=g)=\sum_{w: G(w)=g} P_W(w).$ Applying this to $W=(I, R)$ and $G=(I, F)$,
 a simple calculation gives the result as $(I, F)=(i, f)$ is $I=i\wedge F=f$. $\hfill\square$

\vspace{0.10in} \noindent {\bf Fact 2. } {\em Let $B', B, R$ be  independent random variables with   $B', B$   identically  distributed. Let $G$ be a fixed boolean function.
Then,
\begin{align*}
\Pr(G(R, B)\wedge G(R, B'))=\sum_{r} {P}_R(r)\cdot {\Pr}^2(G(r, B)).
\end{align*}
 }

 \noindent {\bf Proof. } Notice $\Pr(X=x)=\sum_r \Pr(R=r, X=x)$ for variable $R, X.$ Together with Fact 1, we have
\begin{align*}
&\Pr(G(R, B)\wedge G(R, B'))\\
=&\sum_r \Pr(R=r, \{G(R, B)\wedge G(R, B')\}=1)\\
=&\sum_r \Pr(R=r, \{G(r, B)\wedge G(r, B')\}=1)\\
=&\sum_r {P}_R(r)\cdot \Pr(G(r, B))\cdot \Pr(G(r, B'))\\
=&\sum_r {P}_R(r)\cdot {\Pr}^2(G(r, B)),
\end{align*}
where the third equality  uses the independence of $R, B, B'$ and the last equality uses the fact that $B'$ and $B$ are identically distributed. $\hfill\square$

\vspace{.05in}  Now we are ready to present our forking lemma.
\begin{lemma} Let $q\ge 2$ be a fixed integer and $H$ be a set of size $N\ge 2$.  Let $\textsf{A}$ be a randomized algorithm that
 on input $x, h_1, \cdots, h_q$ returns a triple, the first two  elements of which are integers from $\{0, 1, \cdots, q\}$
and the last element of which is a side output. Let $\textsf{IG}$ be a randomized algorithm (called input generator).
The accepting probability of $\textsf{A}$, denoted by $acc$, is defined as the probability that $I, J\ge  1$ in the experiment
\begin{align*}
&x\leftarrow \textsf{IG};\  h_1, \cdots, h_q\leftarrow H; \\
&(I, J, \sigma)\leftarrow \textsf{A}(x, h\lb 1, \cdots, q\rb).
\end{align*}

\noindent The forking algorithm $F_{\textsf{A}}$ associated with $\textsf{A}$ is a randomized algorithm that takes $x$ as input and proceeds as in Fig.   \ref{fig: fork}.

Let $frk=\Pr[F_{\textsf{A}}(x)\ne \textsf{Fail}: x\leftarrow \textsf{IG}].$ Then,
\begin{equation}
frk\ge \frac{8\cdot acc^4}{q^3(q-1)^3}-\frac{3}{N}.
\end{equation}
\end{lemma}
\noindent {\bf Proof. } With respect to $\textsf{Flag}_1$, we define $\textsf{Flag}_1^*$ as event
$$(I_0=\cdots=I_3\ge 1) \wedge (J_0=\cdots=J_3\ge 1)\wedge (J_0>I_0).$$  Then, it is easy to check that $F_A(x)\ne \textsf{Fail}$ is equivalent to $\textsf{Flag}_1^*\wedge \textsf{Flag}_2=1.$ Since $h_{I_0}=\bar{h}_{I_0}$ (resp. $h_{J_0}=\hat{h}_{J_0}$, or,  $\bar{h}_{J_0}=\underline{h}_{J_0}$) in $\neg\textsf{Flag}_2$  holds with probability $1/N.$ It follows that
\begin{align}
\nonumber
frk=&\Pr(\textsf{Flag}_1^*\wedge \textsf{Flag}_2=1)\\
\ge& \Pr(\textsf{Flag}_1^*=1)-3/N. \label{eq: fork-1}
\end{align}
  Notice that
\begin{align}
\nonumber
 &\Pr(\textsf{Flag}_1^*=1)\\
=&\sum_{i=1}^q\sum_{j=i+1}^q \Pr(\wedge_{b=0}^3\{I_b=i\ \wedge \ J_b=j\}). \label{eq: fork-2}
\end{align}

Let $\textsf{A}_1$ (resp. $\textsf{A}_2$, $\textsf{A}_{12}$) be three variants of algorithm $\textsf{A}$ with the only difference in the output which is the first element (resp. the second element, the first two elements) of  $\textsf{A}$'s  output. For instance,
\begin{align}
J_1=&\textsf{A}_2(x, h\lb 1, \cdots, J_0-1, \widehat{J_0, \cdots, q}\rb; \rho), \\
I_2=&\textsf{A}_1(x, h\lb 1, \cdots, I_0-1, \overline{I_0, \cdots, q}\rb;  \rho).
\end{align}
Assigning $I_0=i$ and $J_0=j$, we denote
\begin{align}
  J'_1=&\textsf{A}_2(x, h\lb 1, \cdots, j-1, \widehat{j, \cdots, q}\rb; \rho),\\
I'_2=&\textsf{A}_1(x, h\lb 1, \cdots, i-1, \overline{i, \cdots, q}\rb;   \rho).
\end{align}
 We can similarly define $I_1', J_2',  I_3', J_3'.$ So $I_b, J_b$ for $b\ge 1$ are functions (of \textsf{A}'s inputs and randomness) and when assigning  $I_0=i$ and $J_0=j$, they become $I_b', J_b'$. Hence,
we can apply fact 1 to evaluate Eq. (\ref{eq: fork-2}).   Then, assigning $I_0=i$ and $J_0=j$, applying Fact 1 to realize $I_0=i$ and $J_0=j$ in $\textsf{A}_1$ (for $I_b$) and $\textsf{A}_2$ (for $J_b$), $I_b$ and $J_b$ respectively become $I_b'$ and $J_b'$. Hence,  we have
\begin{align}
&\Pr(\textsf{Flag}_1^*=1)\\
=&\sum_{i=1}^q\sum_{j=i+1}^q \Pr(\wedge_{b=0}^3\{I'_b=i\ \wedge \ J'_b=j\}),  \label{eq: fork-3}
\end{align}
where $I_0, J_0$ is rewritten as $I_0', J_0'$ for brevity (so the term $\{I_0=i\ \wedge \ J_0=j\}$ becomes $\{I'_0=i\ \wedge \ J'_0=j\}$).
Notice $\wedge_{b=0}^1(I_b'=i\wedge J'_b=j)$ is a random variable, with randomness  $R=(x, \rho, h_1, \cdots, h_{i-1})$ and $B=(h_{i}, \cdots, h_q, \hat{h}_{j}, \cdots, \hat{h}_q)$. So we can define
\begin{equation}
\wedge_{b=0}^1(I_b'=i\wedge J'_b=j)=G(R, B)
\end{equation}
 for some boolean function $G$.

Besides, by verifying the definition of $I_b', J_b'$, we can see that
\begin{equation}
\wedge_{b=2}^3(I_b'=i\wedge J'_b=j)=G(R, B')
\end{equation}
 with $B'=(\bar{h}_{i},\cdots, \bar{h}_q,  \underline{h}_{j}, \cdots, \underline{h}_q).$

Hence, applying Fact 2 to Eq. (\ref{eq: fork-3}), we have

\begin{align}
\nonumber
&\Pr(\textsf{Flag}_1^*=1)\\
\nonumber
=&\sum_{\stackrel{r}{1\le i<j\le q}}  {P}_R(r) {\Pr}^2(\wedge_{b=0}^1(I'_{br}=i\ \wedge \ J'_{br}=j))\\
=&\sum_{\stackrel{r}{1\le i<j\le q}}  {P}_R(r) {\Pr}^2(\wedge_{b=0}^1 (I'_{br}, J'_{br})=(i, j)). \label{eq: fork-4}
\end{align}
where $I'_{br}$ (resp. $J'_{br}$) is $I_{b}'$ (resp. $J'_b$) with $R=r.$

Notice that $(I'_{0r}, J'_{0r})=(i, j)$ is a boolean random variable (i.e., the result is true only if the equality  holds),  determined by $h_i, \cdots, h_q$. We can define
\begin{equation}
G'(S, C)\stackrel{def}{=} \{(I'_{0r}, J'_{0r})=(i, j)\}
\end{equation}
for some function $G'$, where  $S=h_{i}, \cdots, h_{j-1}$ and $C=h_j, \cdots, h_q$.

Checking the definition of $(I_{1r}', J_{1r}')$, we can see
\begin{equation}
\{(I'_{1r}, J'_{1r})=(i, j)\}=G'(S, C')
\end{equation}
with $C'=\hat{h}_{j}, \cdots, \hat{h}_q$.

Thus, Eq. (\ref{eq: fork-4}) is
\begin{align}
\nonumber
&\Pr(\textsf{Flag}_1^*=1)\\
=&\sum_{{r}}  {P}_R(r) {\Pr}^2\Big{(}G'(S, C)\wedge G'(S, C')\Big{)}.  \label{eq: fork-5}
\end{align}

Hence,  we can apply Fact 2 to Eq. (\ref{eq: fork-5}) and obtain
\begin{align*}
&\Pr(\textsf{Flag}_1^*=1)\\
= &\sum_{\stackrel{r}{1\le i<j\le q}}{P}_R(r) [\sum_{s}{P}_S(s){\Pr}^2((I'_{0rs}, J'_{0rs})=(i, j))]^2\\
\ge & \sum_{{1\le i<j\le q}} [\sum_{r, s}{P}_R(r) {P}_S(s){\Pr}^2((I'_{0rs}, J'_{0rs})=(i, j))]^2 \\
\ge & \sum_{{1\le i<j\le q}} [\sum_{r, s} {P}_R(r) {P}_S(s){\Pr}((I'_{0rs}, J'_{0rs})=(i, j))]^4\\
=& \sum_{1\le i<j\le q}[\Pr((I'_{0}, J'_{0})=(i, j))]^4,\\
\ge  & \left[\sum_{1\le i<j\le q}\Pr((I'_{0}, J'_{0})=(i, j))\right]^4/(q^3(q-1)^3/2^3)
\end{align*}
where $(I'_{0rs}, J'_{0rs})$ is $(I'_{0r}, J'_{0r})$ with $S=s$,  the first  two inequalities follow from Cauchy-Schwarz inequality\footnote{$\sum_i p_i x_i^2\ge (\sum_i p_i x_i)^2$, if $p_i\ge 0$ and $\sum_i p_i=1$}  (the first one is over distribution $P_R(\cdot)$ and the second one is over distribution $P_{R}(\cdot)P_S(\cdot)$);  the last inequality is to apply Cauchy-Schwarz inequality $\sum_{i=1}^n x_i^2\ge (\sum_i x_i)^2/n$ twice by noticing that $y_i^4=(y_i^2)^2$ so that the first time we use $x_i=y_i^2$ for Cauchy-Schwarz inequality.  Finally, notice that $I_0'=I_0$ and $J_0'=J_0$ by definition. Also,
$\sum_{1\le i<j\le q}\Pr((I_{0}, J_{0})=(i, j))$ is exactly $acc$ by definition. It follows that
$ \Pr(\textsf{Flag}_1^*=1)\ge \frac{acc^4}{q^3(q-1)^3/2^3}$.  From Eq. (\ref{eq: fork-1}), we have $frk\ge \frac{8\cdot acc^4}{q^3(q-1)^3}-3/N.$ $\hfill\square$

\section{Model of Multi-Signature}

In this section, we introduce the model of multi-signature. It consists of the  multi-signature definition and the  security formalization.
\subsection{Syntax}

Mult-signature is a signature with a group of signers, where each of them has a public-key and a private key.  They jointly generate  a signature. The interaction between them proceeds in rounds. Signers are pair-wise connected but the channel is not secure.  The signing protocol is to generate a signature so that the successful verification would indicate that all signers have agreed to sign the message. The target is to generate a compact signature that is shorter than concatenating all signers' individual signatures together.

\vspace{.05in} \begin{definition} A {\bf multi-signature} is a tuple of algorithms ({\bf Setup, KeyGen, Sign, Verify}), described as follows.

\vspace{.10in} \noindent{\bf Setup}.  Given security parameter $\lambda$, it generates a system parameter \textsf{param} that  serves  as part of the input for {\bf KeyGen, Sign, Verify} (but for brevity, we omit it).

\vspace{.10in} \noindent {\bf KeyGen. } It takes \textsf{param} as input and outputs for a user a private  key $sk$ and a public-key $pk$.

\vspace{.10in} \noindent {\bf Sign. } Assume $n$ users with public-keys $(pk_1, \cdots, pk_n)$ want to jointly sign a message $M.$ Then, each user $i$ takes its private key $sk_i$ as input and interacts with other signers. Finally, each of them  outputs a signature $\sigma$ (note: this is for simplicity only; in literature, usually a designated leader outputs $\sigma$). Besides, there is a function $F$ that aggregates  $(pk_1, \cdots, pk_n)$ into a compact public-key $\overline{pk}=F(pk_1, \cdots, pk_n).$

\vspace{.10in} \noindent {\bf Verify. } Upon $(\sigma, M)$ with  the aggregated public-key $\overline{pk}=F(pk_1, \cdots, pk_n)$, verifier  takes $\sigma, M$ and $\overline{pk}$ as input, outputs 1 (for \textsf{accept}) or  0 (for \textsf{reject}).

\end{definition}

\vspace{.10in} \noindent {\bf Remark. } The verify algorithm {\em only} uses the aggregated key $\overline{pk}$ to verify the signature. This is important  for   blockchain, where the recipient only uses $\overline{pk}$ as the public-key. Also, the redeem signature only uses  the multi-signature $\sigma$.  It is desired that both $\overline{pk}$ and $\sigma$ are independent of $n$ while no attacker can forge a valid signature w.r.t. this short $\overline{pk}$.  Even though, our definition generally does not  make any  restriction on $\overline{pk}$ and it especially can be $(pk_1, \cdots, pk_n)$.

\subsection{Security Model}
In this section, we introduce the security model \cite{BDN18} of a multi-signature. It formulates the existential unforgeability. Essentially, it says that no attacker can forge a valid signature on a new message as long as  the signing group contains an honest member. Toward this, the attacker can access to a signing oracle and  create fake public-keys at will. The security is defined    through   a game between a challenger $\textsf{CHAL}$ and an attacker ${\cal A}$.

\vspace{.05in} Initially, \textsf{CHAL} runs {\bf Setup}$(1^\lambda)$ to generate system parameter \textsf{param} and executes  ${\bf KeyGen}$ to generate a public-key $pk^*$ and a private key $sk^*$. It then provides $pk^*|\textsf{param}$ to ${\cal A}$ who interacts with \textsf{CHAL} through  signing oracle below.

\vspace{.05in} \noindent {\bf Sign} ${\cal O}_s(PK, M)$. \quad Here $PK$ is a set of {\em distinct} public-keys with $pk^*\in PK$. Upon this query, \textsf{CHAL} represents $pk^*$ and ${\cal A}$ represents
$PK-\{pk^*\}$ to run the signing protocol on message $M$. Finally, ${\cal O}_s$ outputs the multi-signature $\sigma$ (if it succeeds) or $\perp$ (if it fails).

\vspace{.10in} \noindent {\bf Forgery. } Finally,  ${\cal A}$ outputs a signature $\sigma^*$ for a message  $M^*$, w.r.t. a set of  {\em distinct} public-keys $(pk_1^*, \cdots, pk_N^*)$ s.t. $pk^*=pk_i^*$ for some $i$. ${\cal A}$ succeeds if two conditions are met: (a) $\textsf{Verify}(\overline{pk^*}, \sigma^*, M^*)=1$ (where $\overline{pk^*}=F(pk_1^*, \cdots, pk_N^*)$); (b) no query $((pk_1^*, \cdots, pk_N^*), M^*)$ was issued to  ${\cal O}_s$.  Denote a success forgery event by ${\bf succ}$.

\vspace{.05in} Now we can  define the security of a multi-signature.
\begin{definition} A multi-signature scheme $({\bf Setup,  KeyGen, Sign, Verify})$ is \textsf{existentially unforgeable against chosen message attack} (or \textsf{EU-CMA} for short),  if satisfies the correctness and existential unforgeability below.
\begin{itemize}
\item {\bf Correctness. } For  $(sk_1, pk_1), \cdots, (sk_n, pk_n)$ generated by \textsf{KeyGen}, the signature generated by signing algorithm  on a message $M$ will pass the verification, except for a negligible probability.

 \item {\bf Existential Unforgeability. }  For any PPT adversary ${\cal A}$, $\Pr({\bf succ}({\cal A}))$ is negligible.

\end{itemize}
The multi-signature scheme is said $t$-{\bf EU-CMA},  if  it is {\bf EU-CMA} w.r.t. adversary ${\cal A}$ who always restricts the number of signers in each signing query and  the final forgery to be   at most $t$.

\end{definition}

\section{Model of Canonical Linear Identification}

In this section, we introduce a variant model of canonical identification (ID) scheme and extend it with linearity. We label the ID scheme with a parameter $\tau$. This is needed in order to include our lattice-based ID scheme as a realization for our multi-signature compiler.

\begin{figure*}[!ht]
$
\input xy
\xyoption{all}  \xymatrix@R=0.05in@C=1.3in{
\framebox{\mbox{\bf Prover}}(sk, pk|\tau)&\ar@{}[l] \framebox{\mbox{\bf Verifier}}(pk|\tau) \\
\txt<11.5pc>{$(st, \mbox{CMT})\leftarrow P(\textsf{param})$} \ar[r]^-{\mbox{CMT}} &*\txt<11.5pc>{ CH $\leftarrow \Theta$} \\
\txt<11.5pc>{}& \txt<11.5pc>{}\ar[l]_{\mbox{CH}}  \\
\txt<11.5pc>{$\mbox{Rsp}\leftarrow P(st|sk|pk, \mbox{CH})$ }\ar[r]^{\mbox{Rsp}} &
\txt<11.5pc>{$V_\tau(pk, \mbox{CMT}|\mbox{CH}|\mbox{Rsp})\stackrel{?}{=}1$}
}
$
 \caption{Canonical Identification Protocol}
\label{fig: ID}
\end{figure*}
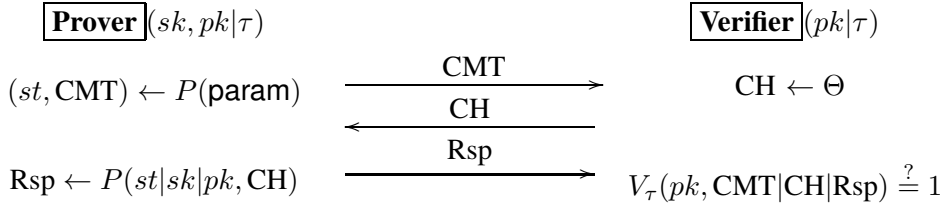

\vspace{.05in} \begin{definition} A canonical identification scheme with parameter $\tau\in \mathbb{N}$ is a  tuple of algorithms ${\cal ID}=({\bf Setup, KeyGen}, P, V_\tau, \Theta)$, where  {\bf Setup} takes security parameter $\lambda$ as input and generates a system parameter $\textsf{param}$;  {\bf KeyGen} is a key generation algorithm that takes $\textsf{param}$  as input and outputs a public key $pk$ and a private key $sk$;   $P$ is an algorithm, executed by {\em prover};  $V_\tau$ is a verification algorithm parameterized by $\tau$, executed by {\em Verifier};  $\Theta$ is a set.   ${\cal ID}$ scheme is a three-round protocol depicted  in Fig. \ref{fig: ID}, where Prover first generates a committing  message $\mbox{CMT}$ with $H_\infty(\mbox{CMT})=\omega(\log \lambda)$, and then Verifier replies with a challenge $\mbox{CH}\leftarrow {\Theta}$ and finally Prover finishes with a response $\mbox{Rsp}$ which will be either rejected or accepted by $V_\tau$.   \label{def: ID-1}
\end{definition}

\vspace{.00in} Denote the domain of $sk$,  $pk$,  \mbox{CMT}, \mbox{Rsp} respectively by   ${\cal SK}, {\cal PK, CMT, RSP}.$  In the following, we define   linearity and  simutability for an ID scheme. Simulatbility appeared before (e.g., \cite{AF+12}) while the linearity is new.

\vspace{.10in}  \noindent {\em Linearity. }  A canonical ID scheme ${\cal ID}=({\bf Setup, KeyGen}, P, V_\tau, \Theta)$  is {\bf linear} if it satisfies the following conditions.
\begin{itemize}
\item[i.] ${\cal SK, PK, CMT, RSP}$ are ${\cal R}$-modules for some ring ${\cal R}$ with $\Theta\subseteq {\cal R}$ (as a set);
 \item[ii.] For any $\lambda_1, \cdots, \lambda_t\in \Theta$ and  public/private pairs $(sk_i, pk_i)$ ($i=1, \cdots, t$), we have that  $\overline{sk}=\sum_{i=1}^t \lambda_i\bullet sk_i$ is a private key of $\overline{pk}=\sum_{i=1}^t \lambda_i\bullet pk_i$.

{\bf Note: } Operator  $\bullet$ between ${\cal R}$ and ${\cal SK}$ (resp. ${\cal PK}, {\cal CMT}, {\cal RSP}$)  might be different  (as long as it is clear from the context), even though   we use the same symbol $\bullet$.

\remove{\item[iii.] There exists an algorithm $P_0$ so that for any $\eta\in \Theta$, $(st, \mbox{CMT})\leftarrow P(\textsf{param})$ and $\mbox{CH}\leftarrow \Theta$, $\mbox{Rsp}=P_0(st, \eta, \mbox{CH})$ is identical to $\mbox{Rsp}$ determined by $sk|pk|\mbox{CH}|(\eta\bullet \mbox{CMT})$.}

\item[iii.] Let  $\lambda_i \leftarrow \Theta$ and $(pk_i, sk_i)\leftarrow \textbf{KeyGen}(1^\kappa),$ for $i=1, \cdots, t.$
If  $\mbox{CMT}_i|\mbox{CH}|\mbox{Rsp}_i$ is a {\em faithfully} generated transcript of the {ID} scheme  w.r.t. $pk_i$,   then
\begin{equation}
V_\tau(\overline{pk}, \overline{\mbox{CMT}}|\mbox{CH}|\overline{\mbox{Rsp}})=1,
\end{equation}
where $\overline{pk}=\sum_{i=1}^t\lambda_i \bullet pk_i, \overline{\mbox{CMT}}=\sum_{i=1}^t \lambda_i\bullet \mbox{CMT}_i$ and $\overline{\mbox{Rsp}}=\sum_{i=1}^t \lambda_i\bullet\mbox{Rsp}_i$.

\end{itemize}

\vspace{.05in} \noindent {\em Simulability.} ${\cal ID}$ is  \textsf{simulatable}  if there exists a PPT algorithm {\bf SIM} s.t. for $(sk, pk)\leftarrow {\bf KeyGen}(1^\lambda)$,  $\mbox{CH}\leftarrow \Theta$ and $(\mbox{CMT}, \mbox{Rsp})\leftarrow {\bf SIM}(\mbox{CH}, pk, \textsf{param})$, it holds that $\mbox{CMT}|\mbox{CH}|\mbox{Rsp}$ is indistinguishable from  a real transcript, even if  the distinguisher  is given   $pk|\textsf{param}$ and has   access to oracle ${\cal O}_{id}(sk, pk)$, where  ${\cal O}_{id}(sk, pk)$ is as  follows:
$(st, \mbox{CMT})\leftarrow P(\textsf{param})$; $\mbox{CH}\leftarrow \Theta$; $\mbox{Rsp}\leftarrow P(st|sk|pk, \mbox{CH})$;  output $\mbox{CMT}|\mbox{CH}|\mbox{Rsp}.$

\vspace{.10in} Now we define the security for an ID scheme. Essentially, it is desired that  an attacker is  unable to impersonate a prover w.r.t. an aggregated public-key, where at least one of the participating public-keys is not generated by attacker. Later we will use this definition to convert an ID scheme into a secure multi-signature.  In our definition,  the prover does not access to  additional information. He is not given extra   capability, either. Thus, our  definition is rather  weak.

\vspace{.05in} \begin{definition}   \label{def: IDsec}
A canonical identification scheme ${\cal ID}=({\bf Setup,  KeyGen}, P, V_\tau, \Theta)$ with linearity and $\tau\in\mathbb{N}$ is {\bf secure}  if it satisfies correctness and security below.

\vspace{.05in} \noindent {\em Correctness. } When no attack presents, Prover will convince Verifier, except for a negligible probability.

\vspace{.05in} \noindent {\em Security.} For any PPT adversary ${\cal A}$,
$\Pr(\textsc{EXP}_{{\cal ID}, {\cal A}}=1)$ is negligible,  where $\textsc{Exp}_{{\cal ID}, {\cal A}}$ is defined as follows, where $pk_i\in {\cal PK}$ for $i\in [t]$ and $\overline{pk}=\sum_{i=1}^t \lambda_i\bullet pk_i$.

\vspace{.05in} \noindent {\bf  Experiment} $\textsc{Exp}_{{\cal ID}, {\cal A}}(\lambda)$

\textsf{param}$\leftarrow {\bf Setup}(1^\lambda)$;

$(pk_1, sk_1)\leftarrow {\bf KeyGen}(\textsf{param})$;

$(st_0, pk_2, \cdots, pk_{t})\leftarrow {\cal A}(\textsf{param}, pk_1)$

$\lambda_1, \cdots, \lambda_t\leftarrow \Theta$

$st_1|\mbox{CMT}\leftarrow {\cal A}(st_0, \lambda_1, \cdots, \lambda_t)$;

 $\mbox{CH}\leftarrow \Theta$; \  $\mbox{Rsp}\leftarrow {\cal A}(st_1, \mbox{CH})$;

$b\leftarrow V_t(\overline{pk}, \mbox{CMT}|\mbox{CH}|\mbox{Rsp})$;

output $b.$

\vspace{.05in}\noindent  ${\cal ID}$ is said {\bf $t^*$-secure} if the security holds for any $t\le t^*.$

\end{definition}

\section{From canonical Linear ID Scheme to Key-and-Signature Compact Multi-signature}

In this section, we show how to convert a linear ID scheme into a multi-sinagure so that the aggregated public-key and signature are both compact.  The idea is to linearly add the member  signatures (resp. public-keys) together with weights  while the weight depends on all public-keys and is different for each user.

\subsection{Construction}
Let $${\cal ID}=({\bf Setup}_{id}, {\bf KeyGen}_{id}, P, V_\tau, \Theta)$$ be a canonical linear ID with  parameter $\tau\in \mathbb{N}$.   $H_0, H_1$ are two random oracles from $\{0, 1\}^*$ to  $\Theta$  with $\Theta\subseteq {\cal R}$, where ${\cal R}$ is the ring defined for the linearity property of ${\cal ID}$.  Our  multi-signature scheme $\Pi=({\bf Setup, KeyGen, Sign, Verify})$ is as  follows.

\vspace{.10in} \noindent {\bf Setup.} \quad Sample and output  $\textsf{param}\leftarrow {\bf Setup}_{id}(1^\lambda)$.
{\em Note: } $\textsf{param}$ should be part of the input to the   algorithms below. But for brevity, we omit it in the future.

\vspace{.10in} \noindent {\bf KeyGen}.  \quad  Sample $(pk, sk)\leftarrow {\bf KeyGen}_{id}(\textsf{param})$;  output a  public-key $pk$ and private key $sk$.

\vspace{.10in} \noindent {\bf Sign.} \quad Suppose that users with public-keys $pk_i, i=1, \cdots, t$ want to jointly sign a message $M$. Let $\lambda_i=H_0(pk_i, PK)$ and $\overline{pk}=\sum_{i=1}^t \lambda_i\bullet pk_i$, where $PK=\{pk_1, \cdots, pk_t\}.$ They run the following procedure.
\begin{itemize}
\item {\em R-1. } \quad User $i$ takes  $(st_i, \mbox{CMT}_i)\leftarrow P(\textsf{param})$ and sends $r_i:=H_0(\mbox{CMT}_i|pk_i)$ to other users.

\item {\em R-2. } \quad  Upon $r_j$ for all $j$ (we do not restrict $j\ne i$ for simplicity), user $i$ verifies if $r_j=H_0(\mbox{CMT}_j|pk_j)$. If no, it aborts; otherwise, it sends $\mbox{CMT}_i$ to other  users.

\item {\em R-3. } \quad Upon $\mbox{CMT}_j, j=1, \cdots, t$, user $i$ computes  $\omx{CMT}=\sum_{j=1}^t \lambda_j \bullet \mbox{CMT}_j$. It  computes $\mbox{CH}=H_1(\overline{pk}|\omx{CMT}|M)$. Finally, it computes $\mbox{Rsp}_i= P(st_i|sk_i|pk_i, \mbox{CH})$ and sends it to other signers.

 \item {\em Output. } Upon $\mbox{Rsp}_j, j=1, \cdots, t$, user $i$ computes $\overline{\mbox{Rsp}}=\sum_{j=1}^t \lambda_j \bullet \mbox{Rsp}_j$, and  outputs the aggregated  public-key $\overline{pk}|t$ and multi-signature $\omx{\mbox{CMT}}|\overline{\mbox{Rsp}}$.

\end{itemize}

\vspace{.10in} \noindent{\bf Verify. } \quad Upon signature $(\omx{CMT}, \overline{\mbox{Rsp}})$ on message $M$ with the aggregated  public key $\overline{pk}|t$, it outputs
$V_t(\overline{pk}, \omx{CMT}|\mbox{CH}|\omx{Rsp})$, where $\mbox{CH}=H_1(\overline{pk}|\omx{CMT}|M).$

\vspace{.10in} \noindent {\bf Remark. }  (1) Since $\overline{pk}|t$ is the aggregated public-key, we assume that it will be correctly computed and available to verifier, which is true for the Bitcoin application. \\
(2)  The most damaging attack to a multi-signature  is the rogue key attack, where an attacker chooses his public-key after seeing other signers' public-keys. By doing this, the attacker could manage to reach an aggregated key for which he knows the private key. In our construction, attacker can not achieve this. Indeed, notice  that $\overline{pk}=H_0(pk_n, PK)\bullet pk_n+\sum_{i=1}^{n-1} H_0(pk_i, PK)\bullet pk_i$, where  $PK=\{pk_1, \cdots, pk_n\}$. The hash-value weights can be computed only after $PK$ has been  determined. Also, if $pk_n$ is the honest user's key, then it is  quite random. So,  $H_0(pk_n, PK)$ (hence $H_0(pk_n, PK)\bullet pk_n$ and also $\overline{pk}$) will be random, given other variables in $\overline{pk}$.  So it is unlikely that attacker can predetermined $\overline{pk}$ and so  the  rogue key attack can not succeed.

\subsection{Security Theorem}

In this section, we prove the security of our scheme. The idea is as follows. We notice that the multi-signature is $(\omx{CMT},\omx{Rsp})$ that satisfies $V_t(\overline{pk}, \omx{CMT}|\mbox{CH}|\omx{Rsp})=1$, where $\mbox{CH}=H_0(\overline{pk}|\omx{CMT}|M)$. Assume $PK=\{pk_1, \cdots, pk_t\}$, where $pk_1$ is an honest user's key and other keys are created by attacker. We want to reduce the multi-signature  security to the security of ID scheme. In this case, $\overline{pk}$ will be the aggregated key with weights  $\lambda_i=H_0(pk_i, PK).$ If an attacker can forge a multi-signature with respect to $\overline{pk}, $ we want to convert it  into an impersonate attack to the ID scheme w.r.t. $\overline{pk}$. There are two difficulties for this task. First,  we need  to simulate the signing oracle without $sk_1$, where we have  to compute the response \mbox{Rsp} for user of $pk_1$ without $sk_1$. Our idea is to  use  the simulability of the ID scheme to help: take a random $\mbox{CH}$ and simulate an ID transcript $\mbox{CMT}'|\mbox{CH}|\mbox{Rsp}'$. Then,  we send $\mbox{CMT}_1=\mbox{CMT}'$ as the committing message. The simulation will be well done if  we can manage to define \mbox{CH} as $H_1(\overline{pk}|\omx{CMT}|M).$ This will be fine if $\overline{pk}|\omx{CMT}|M$ was never  queried to $H_1$ oracle. Fortunately, this is true with high probability: due to the initial registration message at round {\em R-1}, attacker can not know \mbox{CMT}$_1$ before registering  \mbox{CMT}$_j$ using $r_j$ (hence $\mbox{CMT}_j$ is known to us through oracle $H_0$). Hence, $\omx{CMT}$ will have a min-entropy of $H_{\infty}(\mbox{CMT}_1)$, which is super-logarithmic and hence can not be guessed. That is, $\overline{pk}|\omx{CMT}|M$ was unlikely to be  queried to $H_1$ before. Hence, the signing oracle will be simulated without difficulty. The second difficulty is how to convert the forgery into an impersonating attack.  In the ID attack, \mbox{CH} is provided by challenger while in the forgery, \mbox{CH} is the hash  value from $H_1$. The problem is the attacker could make a query $\overline{pk}|\omx{CMT}|M$ to $H_1$ oracle (we maintain) while we do not know whether  this query is toward his final forgery output or not and so we do not know which \mbox{CMT} should be sent to our challenger and consequently we do not know which of such  queries  should be  answered with our  challenger's $\mbox{CH}$. Fortunately, this is not a big issue as we can guess which query will be used for the forgery. There are a polynomial number of such queries. Our random guess only  degrades the success probability by  a polynomial fraction.  This completes our idea. Now we give full details below.

\vspace{.10in} \begin{theorem} Assume that  $h\leftarrow \Theta$ is invertible in ${\cal R}$ with probability $1-\textsf{negl}(\lambda)$.  Let ${\cal ID}=({\bf Setup}_{id}, {\bf KeyGen}_{id}, P, V_\tau, \Theta)$ be  a secure identification scheme with linearity and  simulability.  Then,  our  multi-signature scheme is {\bf EU-CMA} secure. \label{thm: ID2Msg}
\end{theorem}

\vspace{.05in} \noindent {\bf Proof. }
We show that if the multi-signature is broken by ${\cal D}$ with non-negligible probability $\epsilon$, then we can construct an attacker ${\cal B}$ to break  ${\cal ID}$ scheme with a non-negligible probability $\epsilon'$. Given the challenge public-key $pk^*_1$, ${\cal B}$ needs to come up with some other public-keys $pk_2^*, \cdots, pk_\nu^*$ for some $\nu$ of his choice and receives a list of random numbers $\lambda^*_i\leftarrow \Theta$ for $i=1, \cdots, \nu.$ Then, he needs to play as a prover in the ${\cal ID}$ protocol for public-key $\overline{pk^*}=\sum_{i=1}^\nu \lambda^*_i\bullet pk_i^*$ to convince the verifier (his challenger). Toward this, ${\cal B}$ will simulate an environment for ${\cal D}$ and use the responses from ${\cal D}$ to help complete his attack activity. The details follow.

  Upon receiving the challenge public-key $pk^*_1$ and system parameter $\textsf{param}$, ${\cal B}$ samples $\ell^*_{H_0}\leftarrow \{1, \cdots, q^*_{H_0}\}$, where $q^*_{H_0}$ is  the upper bound on the number of {\em new}  queries (i.e., not queried before)  of form $(pk, PK)$ to random oracle $H_0$ s.t. $pk, pk_1^*\in PK$ (call it a \textbf{\em Type-I irregular query}). In addition, a {\em new}  query of format $\mbox{CMT}|\overline{pk^*}|*$ to oracle $H_1$ {\em after} the $\ell_{H_0}^*$th Type-I irregular query will be called a \textbf{\em Type-II irregular query}, where $\mbox{CMT}\in {\cal CMT}$,
  $\overline{pk^*}=\sum_{i=1}^\nu H_0(pk_i^*, PK^*)\bullet pk_i^*$ and $PK^*=\{pk_1^*, \cdots, pk_\nu^*\}$ is the public-key set for the $\ell^*_{H_0}$th  Type-I irregular query.   Let $q^*_{ch}$ be the upper bound on the number of the Type-II irregular queries.
  It then samples $\ell^*_{ch}\leftarrow \{1, \cdots, q^*_{ch}\}.$    ${\cal B}$ invokes ${\cal D}$ with $pk_1^*$ and $\textsf{param}$  and answers his random oracle queries and signing queries as follows.

\vspace{.05in} \noindent {\bf Random Oracle} $H(\cdot)$. \quad For simplicity, we maintain one random oracle $H$ with $H_0(x)=H(0, x)$ and $H_1(x)=H(1, x)$. The query $x$ to $H_b$ is automatically interpreted as query $b|x$ to $H$.  With this in mind, it maintains a hash list ${L}_H$ (initially empty), consisting of records of form $(u, y)$, where $y=H(u)$. Upon a query $b|x$, it first checks if there  was a record $(b|x, y)$ in $L_H$ for some $y$. If yes, it returns $y$; otherwise, there are three cases (all irregular queries will be in these cases as they are unrecorded by definition).

\vspace{.05in} \noindent \textsf{\textbullet\quad  $x$ is not a (Type-I or Type-II) irregular query to $H_b$. }  \quad In this case, it takes $y\leftarrow \Theta$ and adds $(b|x, y)$ into $L_H.$

\vspace{.05in} \noindent \textsf{\textbullet\quad  $x$ is a Type-I irregular query to $H_b$ (thus $b=0$). } \quad In this setting, there are two cases.
\begin{itemize}
\item[-] {\em $x$ is not the $\ell_{H_0}^*$th irregular query. } \quad   In this case,  for each  $pk'\in PK$, it takes $h\leftarrow \Theta$ and  adds $(0|(pk', PK), h)$ into $L_H$. Note for convenience, we treat   each new record in $L_H$ as created due to a hash query (from either simulator ${\cal B}$ or ${\cal D}$). For the technical reason, for given $PK$ with $pk_1^*\in PK$, we treat   $(0|(pk_1^*, PK), *)$ as the last record created in $L_H$ among  all records of $(0|(pk', PK), *)$ with $pk'\in PK$.   Our treatment  is well-defined and perfectly consistent with random oracle, as by our convention, all records of $(pk', PK)$ with $pk', pk^*_1\in PK$ will be recorded in $L_H$ simultaneously whenever it receives a Type-I irregular query (which is $0|x$ in our case).
\item[-] {\em $x$ is the $\ell_{H_0}^*$th irregular query. } \quad  In this case, let $0|x=0|(pk, PK^*)$ with $PK^*=\{pk_1^*, \cdots, pk_\nu^*\}$ for some $\nu\ge 2$. ${\cal B}$ sends $\{pk_2^*, \cdots, pk_\nu^*\}$ to his challenger and receives $\lambda_1^*, \cdots, \lambda_\nu^*$ (each of which is uniformly random over $\Theta$).  Then, ${\cal B}$ inserts $(0|(pk_i^*, PK^*), \lambda_i^*)$ into $L_H$ for $i=1, \cdots, \nu.$ This treatment is perfectly consistent with random oracles:  a Type-I irregular query by definition is an {\em unrecorded} query (i.e.,  not queried before) and $0|(pk', PK^*)$ for each $pk'\in PK^*$ will be recorded in $L_H$ within  one hash  query (thus none of them was queried before).
\end{itemize}

\noindent \textsf{\textbullet\quad  $x$ is a Type-II irregular query to $H_b$ (thus $b=1$). } \quad In this setting, there are two cases.
\begin{itemize}
\item[-] {\em $x$ is not the $\ell_{ch}^*$th Type-II irregular query. } In this case, it takes $y\leftarrow \Theta$ and adds $(1|x, y)$ into $L_H$.

\item[-] {\em $x$ is  the $\ell_{ch}^*$th Type-II irregular query. } In this case, it parses $x=\mbox{CMT}^*|\overline{pk^*}|M^*$ with $\mbox{CMT}^*\in {\cal CMT}.$ Then, it sends $\mbox{CMT}^*$ to its challenger and receive $\mbox{CH}^*$. Then, it adds $(1|x, \mbox{CH}^*)$ to $L_H$.

\end{itemize}

\vspace{.05in} \noindent After our treatment above, $x$ now has been  recorded in $L_H$. Then, the oracle  returns $y$ for $(b|x, y)\in L_H$.

\vspace{.1in}\noindent   {\bf Sign} ${\cal O}_s$ ($pk_1, \cdots, pk_n, M$).  \quad \quad By our security model, it is assumed that  $pk^*_1=pk_t$ for some $t$. Then,  ${\cal B}$ plays the role of user  $pk_t$ while ${\cal D}$ plays  users of  $pk_j$ for $j\ne t$ in the signing algorithm. The action of ${\cal B}$ is as follows.
\begin{itemize}
\item {\em R-1}. \quad  ${\cal B}$ generates $r_t\leftarrow \Theta$ and sends to other signers (played by ${\cal D}$).
\item {\em R-2}. \quad Upon $\{r_j\}_{j\ne t}$ from ${\cal D}$, ${\cal B}$ first issues hash queries $(pk_i, PK)$ for each $pk_i\in PK$ to compute $\lambda_i=H_0(pk_i, PK)$, where $PK=\{pk_1, \cdots, pk_n\}.$ Then, it computes $\overline{pk}$,  takes $h\leftarrow \Theta$ and  runs ${\bf SIM}(h, pk^*, \textsf{param})$ to simulate an ID transcript $(\mbox{CMT}', h, \mbox{Rsp}')$. Then, he  defines  $\mbox{CMT}_t=\mbox{CMT}'$. He also  adds $(0|\mbox{CMT}_t|pk_t, r_t)$ into $L_H$ (in case $(0|\mbox{CMT}_t|pk_t, *)$ not in $L_H$) and otherwise aborts with $\perp$ (denoted by event \textsf{Bad}$_0$).  Next, for each $j\ne t$, it searches a record $(0|\mbox{CMT}_j|pk_j, r_j)$ in $L_H$ for some $\mbox{CMT}_j$ which results in two cases.

(i) \quad  If $(0|\mbox{CMT}_j|pk_j, r_j)$ for all $j\ne t$ are found in $L_H$, it computes $\omx{CMT}=\sum_{i=1}^n \lambda_i\bullet \mbox{CMT}_i$ and checks whether $(1|\overline{pk}|\omx{CMT}|M, y)\in L_H$ for some $y$. If this $y$ does not exist, it records  $(1|\overline{pk}|\omx{CMT}|M, h)$ into $L_H$ and defines $\mbox{CH}=h$ and sends $\mbox{CMT}_t$ to ${\cal D}$; otherwise (denote this event by $\textsf{Bad}_1$),  ${\cal B}$ aborts with $\perp.$

(ii) \quad If $(0|\mbox{CMT}_{j^*}|pk_{j^*}, r_{j^*})$ does not exist in $L_H$ for some  $j^*,$ it sends \mbox{CMT}$_t$ to ${\cal D}$ (normally). However, we remark  that $\mbox{CMT}_{j^*}$ later in Step {\em R-3} (from $j^*$) satisfies $H_0(\mbox{CMT}_{j^*}|pk_{j^*})=r_{j^*}$ (which will be checked there) only negligibly (so this case will not raise  a simulation difficulty), as the hash value is even undefined yet and hence equals $r_j$ with probability $1/|\Theta|$ only, which we ignore it now.

\item {\em R-3}. \quad Upon $\{\mbox{CMT}_j\}_{j\ne t}$, ${\cal B}$ checks if $H_0(\mbox{CMT}_j|pk_j)=r_j$ for each $j.$  If it does not hold for some $j$, ${\cal O}_s$  outputs $\perp$ (normally); otherwise, it sends $\mbox{Rsp}_t:=\mbox{Rsp}'$ to ${\cal D}$.  We clarify two  events: (1) some $\mbox{CMT}_j$ found in {\em R-2}(i) is different from that  received in the current step. In this case, the check in the current step is consistent with a negligible probability  only as $H$ for two different inputs are independent. (2) {\em R-2}(ii) occurs to some $j^*$ (so \mbox{CMT}$_{j^*}$ is not found there)  while \mbox{CMT}$_{j^*}$ received in the current  step is consistent with $r_j$. As seen above, this holds with probability $1/|\Theta|$ only.  Ignoring these events, $\mbox{CH}$ and $\{\mbox{CMT}_j\}_j$ are determined in {\em R-2}(i) and $\{\mbox{CMT}_j\}_j$ are consistent with those received in the current step.

\item {\em Output. } Upon $\mbox{Rsp}_j$ for $j\ne t$, it computes $\omx{Rsp}=\sum_{j=1}^n \lambda_j\bullet \mbox{Rsp}_j.$ The final signature is $(\omx{CMT}, \omx{Rsp})$ with the aggregated key $\overline{pk}|t.$

\end{itemize}

\vspace{.05in}
Finally, ${\cal D}$ outputs a forgery $(\alpha, \beta)$ for message $M'$ and public keys $PK'$. If $\alpha|PK'|M'\ne \mbox{CMT}^*|PK^*|M^*$ or $\alpha|\beta$ is invalid (when verified using $V(\cdot)$), ${\cal B}$ exits with $\perp$; otherwise, he verifies $(\alpha, \beta)$. If invalid, he  outputs $\perp$; otherwise, he defines  $\mbox{Rsp}^*=\beta$ and sends it back to his challenger.  This completes the description of ${\cal B}$.

We now analyze the success probability of ${\cal B}.$
First, the view of ${\cal D}$ is identical to the real game, except for the following events.
\begin{itemize}
\item[a.] In step {\em R-2} of ${\cal O}_s$, $(\mbox{CMT}', h, \mbox{Rsp}')$ is simulated by {\bf SIM} (instead of being  computed using $sk^*_1$).  However, by hybrid reduction to simulability of ${\cal ID}$, the view of ${\cal D}$ is statistical close from his view when  this transcript is generated using $sk^*_1$ (with the same $h).$
\item[b.] In step {\em R-2} of oracle ${\cal O}_s$, when   $(0|\mbox{CMT}_t|pk_t, y)\in L_H$, \textsf{Bad}$_0$ occurs for some $y$ (hence the view of ${\cal D}$ is inconsistent if $y\ne r_t$). However, since $\mbox{CMT}'$ (i.e., $\mbox{CMT}_t$) is just simulated in this oracle query and $H_\infty(\mbox{CMT}')=\omega(\log\lambda)$, $\mbox{CMT}'$ is independent of current records in $L_H$. Hence, $\textsf{Bad}_0$ occurs with probability at most $Q/2^{H_\infty(\mbox{CMT}')}$ (negligible), where $Q$ is the number of records in $L_H$. We ignore this negligible  probability from now on.
\item[c.] In step {\em R-2} (i), if $(1|\overline{pk}|\omx{CMT}|M, y)\in L_H$ for some $y$, then event \textsf{Bad}$_1$ occurs. In this case, \textsf{A} can not define $\mbox{CH}=h$ and the simulation can not continue. However, since $\omx{CMT}=\lambda_t\bullet \mbox{CMT}'+\sum_{j\ne t}\lambda_j\bullet \mbox{CMT}_j$ and $\mbox{CMT}'$ is simulated in the current oracle and hence independent of the rest variables in this equation. Hence, as long as $\lambda_t$ is invertible (which is violated only negligibly), $\omx{CMT}$ has a min-entropy at least $H_\infty(\mbox{CMT})=\omega(\log\lambda).$ Thus, similar to \textsf{Bad}$_0$ event,  \textsf{Bad}$_1$ occurs negligibly only.
 \item[d.] Finally, when ${\cal D}$ outputs $(\alpha, \beta)$ for message $M'$ and public-key set $PK'$, it has  $\alpha|PK'|M'\ne \mbox{CMT}^*|PK^*|M^*$. Since $(\alpha, \beta)$ has been verified, a Type-I irregular query $(pk, PK')$ and a Type-II irregular query $\alpha|\overline{pk'}|M$ must have been issued (the first query $(pk, PK')$ for some $pk\in PK'$ is the Type-I irregular query while the first query of $\alpha|\overline{pk'}|M$ is the Type-II query;  the existence of such queries are guaranteed as the verification of $(\alpha, \beta)$ by ${\cal B}$ will certainly  issue these queries). Since $\ell_H^*$ and $\ell_{ch}^*$ are chosen uniformly random,  they happened to the foregoing two queries with probability $\frac{1}{q_H^*q^*_{ch}}\ge \frac{1}{q_0q_1},$ where $q_0$ (resp. $q_1$) is  the upper bound on $\sharp$ queries to $H_0$ (resp. $H_1$).

\end{itemize}
From the analysis of (a)(b)(c), their occurrence changes the adversary view negligibly. Ignoring this, from item d, when $\ell^*_H$ and $\ell_{ch}^*$ is chosen correctly, the view of ${\cal D}$ is indistinguishable from its view in the real game. On the other hand, it is easy to verify that conditional on this correct choice, a valid forgery indicates a successful attack by ${\cal B}$. Hence,  ${\cal B}$ can break the ID security with probability at least $\epsilon/q_0q_1,$ non-negligible. This contradicts the security of our ID scheme.  $\hfill\square$

\vspace{.05in} If the adversary always restricts the number of signers in  the signing query and the forgery to be at most $T$, then Theorem immediately implies the following corollary.

\vspace{.10in} \begin{corollary} Let $T\ge 2$. Assume that  $h\leftarrow \Theta$ is invertible in ${\cal R}$ with probability $1-\textsf{negl}(\lambda)$.  Let ${\cal ID}=({\bf Setup}_{id}, {\bf KeyGen}_{id}, P, V_\tau, \Theta)$ be  a $T$-secure identification scheme with linearity and  simulability.  Then,  our  multi-signature scheme is $T$-{\bf EU-CMA} secure. \label{cor: ID2Msg}
\end{corollary}

\section{Realizations}

In this section, we will realize our compiler with ID schemes: Schnorr ID scheme and a lattice-based ID scheme. The first scheme is similar to Boneh et al. \cite{BDN18}. But we keep it as it is very simple and efficient and can demonstrate the usage of our compiler.  The second one is new and breaks a barrier that the previous schemes can not overcome.

\subsection{Realization I: Schnorr Identification}

In this section, we apply our compiler to the well-known Schnorr ID scheme \cite{Schnorr}. Toward this, we only need to show that it is linear with simulability and security.  For clarity, we first review this scheme.

\begin{figure*}[!ht]
$
\input xy
\xyoption{all}  \xymatrix@R=0.05in@C=1.3in{
\framebox{\mbox{\bf Prover}}(s, A=g^s)&\ar@{}[l] \framebox{\mbox{\bf Verifier}}(A=g^s) \\
\txt<11.5pc>{$x\leftarrow \mathbb{Z}_q, X=g^x$} \ar[r]^-{X} &*\txt<11.5pc>{$c\leftarrow \mathbb{Z}_q$} \\
\txt<11.5pc>{}& \txt<11.5pc>{}\ar[l]_{c}  \\
\txt<11.5pc>{$ z=sc+x$ mod $q$}\ar[r]^{z} &
\txt<11.5pc>{$g^z\stackrel{?}{=}A^cX$}
}
$
 \caption{Schnorr  Identification Scheme}
\label{fig: Schnorr}
\end{figure*}
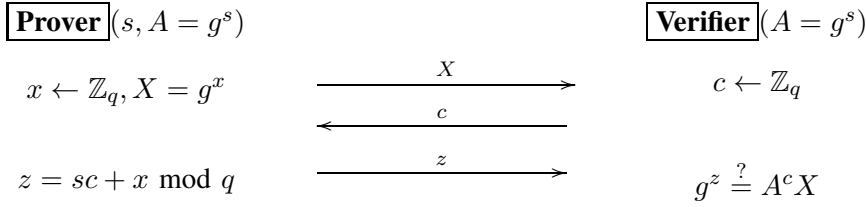

Let $q$ be a large prime.  Consider a prime group of order $q$   with a random generator $g$ (e.g., the group on elliptic curve secp256k1 of  $y^2=x^3+7$ for Bitcoin).    The Schnorr identification is depicted  in Fig. \ref{fig: Schnorr}. This scheme can be regarded as a realization  of the  parameterized ID scheme with  parameter $\tau$ never used.
    In the following, we show that Schnorr ID scheme satisfies the three properties.

\vspace{.10in} \noindent {\bf Linearity. } \quad Notice that ${\cal SK}={\cal RSP}={\cal R}=\Theta=\mathbb{Z}_q$, ${\cal CMT}={\cal PK}=\langle g\rangle$. We now verify the linearity property.
\begin{itemize}

\item[i.] As seen in Section \ref{sect: R-module}, $\mathbb{Z}_q$ and $\langle g\rangle$ are both $\mathbb{Z}_q$-modules, where the multiplication $\bullet$ between ${\cal R}=\mathbb{Z}_q$ and $\mathbb{Z}_q$ is the multiplication of  $\mathbb{Z}_q$,  while $\bullet$ between ${\cal R}=\mathbb{Z}_q$ and $\langle g\rangle$ is exponentiation: $s\bullet m=m^s$.    Hence, ${\cal SK}, {\cal PK}, {\cal CMT}, {\cal RSP}$ are ${\cal R}$-modules.

\item[ii.] Let $pk_i=g^{s_i}$ with $sk_i=s_i, i=1, \cdots, n. $   Let $\lambda_1, \cdots, \lambda_n\in {\cal R}$.  Then, $\overline{sk}=\sum_{i=1}^n \lambda_i\bullet sk_i=
\sum_{i=1}^n \lambda_i s_i$, where the addition is the group operation for  ${\cal SK}$ (i.e., addition in $\mathbb{Z}_q$).  Note the  group operation for ${\cal PK}$ is the multiplication in $\langle g\rangle.$ Hence, $\overline{pk}=\prod_{i=1}^n \lambda_i\bullet pk_i=\prod_{i=1}^n pk_i^{\lambda_i}=g^{\sum_{i=1}^n \lambda_i s_i}.$ Thus, $\overline{sk}\in {\cal SK}$ is the private key of $\overline{pk}\in {\cal PK}.$

\item[iii.] Let $X_i|c|z_i$ be a transcript of ${\cal ID}$ w.r.t., $pk_i=g^{s_i}$ and $sk_i=s_i, i=1, \cdots, n. $ For $\lambda_i\in {\cal R}$, $\overline{X}=\prod_{i=1}^n \lambda_i\bullet X_i=\prod_{i=1}^n X_i^{\lambda_i}$ and  $\overline{z}=\sum_{i=1}^n \lambda_i\bullet z_i=\sum_{i=1}^n \lambda_i z_i.$  If $g^{z_i}=pk_i^cX_i, $ then $\prod_{i=1}^n g^{\lambda_i z_i}=\prod_{i=1}^n (pk_i^cX_i)^{\lambda_i}.$ Hence,
$g^{\overline{z}}=\overline{pk}^c\overline{X}, $  desired!
\end{itemize}

\vspace{.10in} \noindent {\bf Simulability. } \quad Let $pk=g^s$ be the public-key and $sk=s$ be the private key. For $c\leftarrow \mathbb{Z}_q$, we define ${\bf SIM}$ by  taking  $z\leftarrow \mathbb{Z}_q$ and $X=g^{z}pk^{-c}.$ The simulated {\cal ID} transcript is $X|c|z.$ Obviously, this transcript is valid (i.e., it passes the verification).  Now we show that for any (even unbounded) distinguisher ${\cal D}$ that has oracle access to ${\cal O}_{id}$ can not distinguish the output of ${\bf SIM}$ from the real {\cal ID} transcript. Notice for both simulated and real transcripts $X|c|z$, it satisfies $g^z=pk^cX$. Hence, $X=g^x$ for some $x\in \mathbb{Z}_q$ and $z=cs+x.$ In the real transcript, $x\leftarrow \mathbb{Z}_q$ while the simulated transcript $z\leftarrow \mathbb{Z}_q$. Hence, given $c$, $(x, z)$ (hence $X, z$) in both transcripts has the same distribution. Since $c$ is uniformly random in $\mathbb{Z}_q$ in the simulation, the simulated and real transcripts have the same distribution (independent of adversary view before the challenge which includes the responses from ${\cal O}_{id}$). Thus, the adversary view, given oracle access to ${\cal O}_{id}$,  in both cases has the same distribution.  The simulability follows.

\vspace{.10in} \noindent {\bf Security. } \quad  We now prove the security of Schnorr ID scheme under Definition \ref{def: IDsec}.
\begin{lemma} Under discrete logarithm assumption, Schnorr ID scheme is secure w.r.t. Definition \ref{def: IDsec}.  \label{le: DL}
\end{lemma}
\noindent {\bf Proof. } If there exists an adversary ${\cal D}$ that breaks the Schnorr ID scheme with non-negligible probability $\epsilon$, then we construct an adversary ${\cal A}$ that breaks discrete logarithm in $\langle g\rangle$ with a non-negligible probability $\epsilon'$. The idea is to make use of ${\cal D}$ to construct an algorithm $\textsf{A}$ for the nested forking lemma and then use the output of the forking algorithm to derive the discrete logarithm for the challenge. Upon a challenge $A_1=g^x$ and parameters $q, g$, ${\cal A}$ constructs $\textsf{A}((A_1, g, q), \lambda_1, c; \rho)$ as follows (so $h_1|h_2=\lambda_1|c$ with $q=2$ in the forking algorithm), where $\overline{A}=\sum_{i=1}^t A_i^{\lambda_i}$.

\vspace{.05in} \noindent {\bf  Algorithm} $\textsf{A}((A_1, g, q),\lambda_1, c; \rho)$

{\bf Parse} $\rho$ as two parts: $\rho=\rho_0|\rho_1$

$(st_0, A_2, \cdots, A_{t})\leftarrow {\cal D}(q, g, A_1; \rho_0)$

$\lambda_2, \cdots, \lambda_t\leftarrow \mathbb{Z}_q$ using randomness $\rho_1$

$st_1|X\leftarrow {\cal D}(st_0, \lambda_1, \cdots, \lambda_t)$;

 $z\leftarrow {\cal D}(st_1, c)$;

{\bf If} $g^z=\overline{A}^c\cdot X$, {\bf then} $b=1$;

{\bf else} $b=0$;

{\bf output} $(b, 2b, \{A_i|\lambda_i\}_1^t|X|z|c|g|q).$

\vspace{.05in} \noindent From the description of $\textsf{A}$ and the forking algorithm $F_A$ (for the forking lemma),  the rewinding in the forking algorithm $F_A$ only changes $\lambda_1$ and/or $c$ as well as those affected by $(\lambda_1, c)$. In terms of forking lemma terminology, we have $(h_1, h_2)=(\lambda_1, c)$ and $I_0=1, J_0=2$ (for a successful execution; otherwise, $\textsf{A}$ will abort when $I_0\le J_0$).   Let us now analyze algorithm forking algorithm $F_{\textsf{A}}.$ When four executions are executed successfully (i.e., $b=1$ for all cases), then the output for each execution will be described as follows.
Let $A_i=g^{a_i}$ for $i=1, \cdots, t.$
\begin{itemize}
\item[-] {\em Execution 0. } It outputs $(1, 2, \{A_i|\lambda_i\}_1^t|X|z|c|g|q)$.  As the verification passes,
\begin{equation}
z=(\sum_{i=1}^t \lambda_i a_i)c+x,  \label{eq: DL0-0}
\end{equation}
where $X=g^x.$

\item[-] {\em Execution 1. } Compared with execution 0, the input only changes $c$ to $\hat{c}$. From the code of $\textsf{A}$, the output is
$(1, 2, \{A_i|\lambda_i\}_1^t|X|\hat{z}|\hat{c}|g|q)$.
As the verification passes,
\begin{equation}
\hat{z}=(\sum_{i=1}^t \lambda_i a_i)\hat{c}+x. \label{eq: DL0-1}
\end{equation}

\item[-] {\em Execution 2. } Compared with execution 0, the input changes $\lambda_1$ to $\bar{\lambda}_1$ and $c$ to $\bar{c}$. From the code of $\textsf{A}$, the output is
$(1, 2, \{A_i|\lambda_i\}_2^t|A_1|\bar{\lambda}_1|X'|\bar{z}|\bar{c}|g|q)$.
As the verification passes,
\begin{equation}
\bar{z}=(\bar{\lambda}_1a_1+\sum_{i=2}^t \lambda_i a_i)\bar{c}+x',  \label{eq: DL0-2}
\end{equation}
where $X'=g^{x'}.$

\item[-] {\em Execution 3. } Compared with execution 0, the input changes $\lambda_1$ to $\bar{\lambda}_1$ and $c$ to $\underline{c}$. From the code of $\textsf{A}$, the output is $(1, 2, \{A_i|\lambda_i\}_2^t|A_1|\bar{\lambda}_1|X'|\underline{z}|\underline{c}|g|q)$.
    As the verification passes,
\begin{equation}
\underline{z}=(\bar{\lambda}_1a_1+\sum_{i=2}^t \lambda_i a_i)\underline{c}+x'.  \label{eq: DL0-3}
\end{equation}
\end{itemize}
From Eqs. (\ref{eq: DL0-3})(\ref{eq: DL0-2}), ${\cal A}$ can derive $\bar{\lambda}_1a_1+\sum_{i=2}^t \lambda_i a_i$, as long as $c\ne c'$ in $\mathbb{Z}_q$. Similarly, from Eqs. (\ref{eq: DL0-1})(\ref{eq: DL0-0}), ${\cal A}$ can derive ${\lambda}_1a_1+\sum_{i=2}^t \lambda_i a_i$, as long as $\bar{c}\ne \underline{c}.$
This can further give $a_1$, as long as $\lambda_1\ne \bar{\lambda}_1$ in $\mathbb{Z}_q$. Finally, if the forking algorithm does not \textsf{fail}, then the four executions succeeds {\bf and}  $(c\ne c')\wedge (\bar{c}\ne \underline{c})\wedge (\lambda_1\ne \bar{\lambda}_1)=$True. By forking lemma, it does not fail with   probability at least $\epsilon^4/(1\cdot 1)-3/|\Theta|=\epsilon^4-3/q$. Hence, ${\cal A}$ can obtain $a_1$ with probability at least $\epsilon^4-3/q,$ non-negligible. This contradicts to the discrete logarithm assumption. $\hfill\square$

\vspace{.10in} \noindent {\bf Key-and-Signature Compact Multi-Signature from Schnorr {ID} Scheme. }  Since Schnorr ID scheme satisfies the linearity, simulability and special  soundness,  the multi-signature from this scheme using our compiler is obtained.  For clarity, we give the complete signature in the following.  Let $pk_i=g^{s_i}$ be the public-key with private key $sk_i=s_i$ for $i=1, \cdots, n$. When users $PK=\{pk_1, \cdots, pk_n\}$ want to jointly sign a message $M$, they act as follows.
\begin{itemize}
\item[\textbullet] {\bf R-1.} \quad  User $i$ generates $X_i=g^{x_i}$ for $x_i\leftarrow \mathbb{Z}_q$ and sends $H_0(X_i|pk_i)$ to other users.

\item[\textbullet] {\bf R-2.} \quad  Upon $\{r_j\}_{j=1}^n$, user $i$ sends $X_i$ to other users.

\item[\textbullet] {\bf R-3.}  Upon $\{X_j\}_{j=1}^n$, user $i$ checks $r_j\stackrel{?}{=}H_0(X_j|pk_j)$ for all $j$. If not, he rejects; otherwise,  he  computes
\begin{align}
\overline{pk}=&\prod_{i=1}^n pk_i^{H_0(pk_i, PK)}\\
\overline{X}=&\prod_{i=1}^n X_i^{H_0(pk_i, PK)}.
\end{align}
 Then, he computes
\begin{align}
c=H_1(\overline{pk}|\overline{X}|M), \ z_i=s_i c+x_i
\end{align}
and sends $z_i$ to leader.
\item[\textbullet] {\em Output. } \quad Receiving  all $z_j$'s, user $i$ computes
$$\overline{z}=\sum_{j=1}^n H_0(pk_j,PK) z_j.$$  Finally, it outputs $(\overline{X}, \overline{z})$ as the multi-signature of $M$ with the aggregated public-key $\overline{pk}$ (note: the compiler protocol includes $n$ in the aggregated key; we omit it here as it is not used in  the verification).

\item[\textbullet] {\em Verification. } \quad  To verify signature $(\overline{X}, \overline{z})$ for $M$ with the aggregated public-key $\overline{pk}$, it computes $c=H_1(\overline{pk}|\overline{X}|M)$. It accepts only if $g^{\overline{z}}=\overline{pk}^c \cdot \overline{X}.$
\end{itemize}

We denote this signature scheme by \textsf{Schnorr-MultiSig}. Notice that $c\leftarrow \mathbb{Z}_q$ is invertible in ${\cal R}$ with probability $1-1/q$. As it satisfies linearity, simulability and security, by Theorem \ref{thm: ID2Msg}, we have the following.

\vspace{.05in} \begin{corollary} If Discrete logarithm  assumption in  $\langle g\rangle$ holds, then \textsf{Schnorr-MultiSig} is EU-CMA.

\end{corollary}

\vspace{.05in} \noindent {\bf Remark}.  Boneh et al. \cite{BDN18} proposed a method that  transforms Schnorr ID to a  key-and-signature compact multi-signature. Their protocol is an improvement of Maxwell et al. \cite{MP+18} to overcome a simulation flaw.  Their protocol is also 3-round but  computationally more efficient in the signing process than ours. However, our sizes of aggregated (public-key, signature) as well as  the verification cost are all the same as theirs (also identical to  the original Schnorr signature case). Aggregated public-key and signature have impacts  on the storage at a large number of blockchain nodes and the verification cost has the impact on the power consumption on  these nodes. The signing cost is relatively not so important as it only has impact on  the involved signers.   Boneh et al. \cite{BDN18} uses $\lambda_is_i$ as a secret for public-key $pk_i^{\lambda_i}$ to generate a member signature $X_i|c|z_i$ and the final multi-signature $\widetilde{X}=\prod_i X_i$ and $\tilde{z}=\sum_i z_i$. Their main saving (over us) is to  avoid $n$ exponentiations in computing our $\overline{X}$. One  might be motivated to modify our general  compiler so that it  uses $\lambda_i \bullet pk_i$ (whose private key is $\lambda_i\bullet sk_i$) to generate a member signature $\mbox{CMT}_i|\mbox{Rsp}_i$ so that the final multi-signature is $\wmx{CMT}|\wmx{Rsp}$ with $\wmx{CMT}=\sum_i \mbox{CMT}_i$ and $\wmx{Rsp}=\sum_i\mbox{Rsp}_i$. However, this looking secure scheme has a simulation issue in general  when we prove Theorem \ref{thm: ID2Msg}:  it is required that $\{{\bf SIM}(\mbox{CH}, \lambda\bullet pk)\}_\lambda$ is  indistinguishable from the list of real transcripts for a fixed but random $pk$ while it is not clear how this can be proven {\em generally}.

\subsection{Realization II:  a new lattice-based ID scheme}
In this section, we propose a new ID scheme from lattice and then apply our compiler to obtain a lattice-based multi-signature scheme. This is  the first lattice-based multi-signature that has both a compact public-key and a compact signature without    a restart during  the  signing process.

\vspace{.05in} \noindent {\bf Notations. } \quad The following  notations are specific for this section (see Section \ref{sect: Pre} for more).
\begin{itemize}
 \item As a  convention for lattice over ring, this section   uses security parameter $n$ (a power of 2), instead of $\lambda$;
 \item $q$ is a prime with  $q\equiv 3$ mod 8;
 \item $R=\mathbb{Z}[x]/(x^n+1)$; $R_q=\mathbb{Z}_q[x]/(x^n+1)$;  $R_q^*$ is  the set of invertible elements in $R_q$;
  \item for a vector ${\bf w}$, we implicitly assume it is a column vector and the $i$th component is $w_i$ or ${\bf w}[i]$;
    \item for a matrix or vector $X$, $X^T$ is its transpose;
   \item ${\bf 1}$ denotes the all-1 vector $(1, \cdots, 1)^T$ of dimension that is  clear  from  the context;
   \item for $u=\sum_{i=0}^{n-1} u_ix^i\in R$, $||u||_\infty=\max_i |u_i|;$ 
   \item $\alpha\in \mathbb{Z}_q$ always uses   the default representative with  $-(q-1)/2\le \alpha\le (q-1)/2$ and similarly, for $u\in R_q$,  each coefficient of $u$ by default belongs to this range;
  \item  $e=2.71828\cdots$ is the Euler's number;
\item ${\cal C}=\{c\in R\mid ||c||_\infty\le \log n, \deg(c)<n/2\}$
\item ${\cal Y}=\{y\in R \mid ||y||_\infty \le n^{1.5}\sigma\log^3 n\}$
\item ${\cal Z}=\{z\in R \mid  ||z||_\infty \le (n-1)n^{1/2}\sigma\log^3 n\}$.
\end{itemize}

\subsubsection{Ring-LWE and Ring-SIS}  In this section, we introduce the ring-LWE amd ring-SIS assumptions (see \cite{LPR13,PR06,LM06} for details). For $\sigma>0$, distribution $D_{\mathbb{Z}^n, \sigma}$ assigns the probability proportional to $e^{-\pi ||{\bf y}||^2/\sigma^2}$ for any ${\bf y}\in \mathbb{Z}^n$ and 0 for other cases. As in \cite{AF+12}, $y\leftarrow D_{R, \sigma}$ samples  $y=\sum_{i=0}^{n-1} y_i x^i$ from $R$ with  $y_i\leftarrow D_{\mathbb{Z}, \sigma}.$

The  \textsf{Ring Learning With Error} (Ring-LWE$_{q, \sigma, 2n}$) problem over $R$ with standard deviation $\sigma$ is defined as follows. Initially, it takes $s\leftarrow D_{R, \sigma}$ as secret. It then takes $a\leftarrow R_q, e\leftarrow D_{R, \sigma}$ and outputs $(a, as+e)$. The problem is to distinguish $(a, as+e)$ from a tuple $(a, b)$ for $a, b\leftarrow R_q.$    The \textsf{Ring-LWE}$_{q, \sigma, 2n}$ assumption is to say that no PPT algorithm can solve Ring-LWE$_{q, \sigma, 2n}$ problem with a non-negligible advantage. According to \cite{LPR10,DD12}, ring-LWE assumption with $\sigma=\tilde{\Omega}(n^{3/4})$ is provably hard and so it is safe to assume $\sigma=\Omega(n)$.

The \textsf{Small Integer Solution problem} with parameters $q, m, \beta$ over ring $R$ (ring-SIS$_{q, m, \beta}$) is as follows: given $m$ uniformly random elements $a_1, \cdots, a_m$ over $R_q$, find $(t_1, \cdots, t_m)$ so that $||t_i||_\infty\le \beta$ and $a_1t_1+\cdots+a_mt_m=0$ ({\bf note}: here we use $||\cdot||_\infty$ norm while the literature regularly uses square-root norm $||\cdot||$. However, the gap  is only a factor $n$ on $\beta$ and does not affect the validity of the assumption according to the current research status for ring-SIS).  We consider the case  $m=3$.  As we use $q=3$ mod 8, by \cite[Theorem 1]{BGM93},  $x^n+1=\Phi_1(x)\Phi_2(x)$ for irreducible polynomials $\Phi_1(x), \Phi_2(x)$ of degree $n/2$. So by Chinese remainder theorem,  $a_i$ is invertible, except for probability $2q^{-n/2}$.  Hence, ring-SIS is equivalent to the case of  invertible $a_2$ which is further equivalent to problem $a_1t_1+t_2+a_3t_3=0$, as we can multiply it by $a_2^{-1}$. By \cite{LM06,CDW17}, the best {\em quantum} polynomial algorithm for  ring-SIS problem with $q, m$ can only solve $\beta=2^{\tilde{O}(\sqrt{n})}$ case. Thus, it is safe to assume Ring-SIS$_{q, m, \beta}$ for any polynomial $\beta$ or  even $\beta=2^{\sqrt[4]{n}}$.


\vspace{.05in} \subsubsection{Construction}

We now describe our new ID scheme from  ring $R$.
Initially, take $s_1, s_2\leftarrow D_{R, \sigma}, a\leftarrow R_q^*$ and compute $u=as_1+s_2$. The system parameter is $a$; the public key is $u$ and the private key is $(s_1, s_2).$  Our ID scheme is as follows; also see Fig. \ref{fig: Ring-LWE}.
\begin{itemize}

\item[1.] Prover  generates ${\bf y}_{1}, {\bf y}_{2} \leftarrow {\cal Y}^\mu$ and computes ${\bf v}=a{\bf y}_{1}+{\bf y}_{2}$ and sends ${\bf v}$ to Verifier, where $\mu\ge \log^2 n.$

\item[2.] Receiver samples $c\leftarrow {\cal C}$ and sends it to Prover.

\item[3.] Upon $c$, Prover does the following:
\begin{itemize}
\item[a.] Compute
${\bf z}_{1}=s_1c\cdot {\bf 1}+{\bf y}_{1},\  {\bf z}_{2}=s_2c\cdot {\bf 1}+{\bf y}_{2};$
\item[b.] Let $A=\{j\mid  z_{1j}, z_{2j}\in {\cal Z}\}.$ If $A=\emptyset$, abort; otherwise, take $j^*\leftarrow A$ and compute
\begin{equation*}
z_1=z_{1j^*}+\sum_{j\ne j^*} y_{1j}, \ z_2=z_{2j^*}+\sum_{j\ne j^*} y_{2j}.
\end{equation*}

\end{itemize}

\item[4.]  Upon $z_1, z_2$, Verifier checks
\begin{equation*}
\sum_{i=1}^\mu v_i\stackrel{?}{=}az_1+z_2-uc, \ ||z_b||_\infty\stackrel{?}{\le} \eta, b=1, 2,
\end{equation*}
where $\eta_t=5\sigma n^2 \sqrt{t\mu}\log^6 n$ and $t$ is a positive integer (see the remark below).   If all are valid, it accepts; otherwise, it rejects.
\end{itemize}

\vspace{.05in}\noindent {\bf Remark. } We give two clarifications.

\noindent (1) \quad The correctness  does not need  $\eta_t$ to vary with $t$ as it is defined so. Actually, $\eta_1 =3\sigma n^{1.5}\sqrt{\mu}\log^4n$ suffices for all $t$. However, we need the dependency on $t$ for the linearity and later for the  multi-signature.  Especially, for linearity with  $t$ transcripts,  $\eta_t$ is needed to depend on $t$. Further, for  the multi-signature scenario, $t$ stands for the number of signers.

\noindent (2) \quad  It should be pointed out that the choice of $j^*$ (if it exists) does not affect $z_1, z_2$ at all as $z_b=s_bc+\sum_{j=1}^t y_{bj}$ for $b=1, 2$. In addition, the probability that $j^*$ does not exist is exponentially small in $n$ and so defining  $j^*$ is unnecessary. However, we keep it for ease of analysis later.

\vspace{.10in} \noindent {\bf Correctness. } We now prove  the correctness with $\eta_t$ replaced by a smaller value  $\eta_1=3 \sigma n^{1.5}\sqrt{\mu}\log^4n$. When all signers are honest, the  protocol is easily seen to be correct if we can show $A=\emptyset$ or $||z_b||_\infty> \eta_1$ has a negligible probability. The former is shown in Lemma \ref{le: A} below.  For the latter, notice that $z_1=s_1c+y_{11}+\cdots+y_{1\mu}.$ If we use $\underline{w}\in R$ to denote the coefficient vector of the polynomial $w$. Then,
\begin{align}
\underline{y_{11}+\cdots+y_{1\mu}}=\underline{y_{11}}+\cdots+\underline{y_{1\mu}}. \label{eq: correct}
\end{align}
Notice each component of $\underline{y_{1j}}$ is uniformly random in $\{-\sigma n^{1.5}\log^3n, \cdots, \sigma n^{1.5}\log^3n\}.$ By Hoeffding inequality on each of the vector component  in Eq. (\ref{eq: correct}),
$||\sum_i \underline{y_{1i}}||_\infty>2\sigma n^{1.5}\sqrt{\mu}\log^4n$ only has a probability at most $2ne^{-\log^2n}.$ By Lemma \ref{le: sc} below, $||sc||_\infty>\sigma n^{1/2}\log^3n$ with probability at most $e^{-\Omega(\log^2n)}.$ Hence, correctness holds for bound $\eta_1$,  except for probability at most $e^{-\Omega(\log^2 n)}$ ({\bf note}: for brevity, this quantity should be understood as there exists constant $C$ so that the exception probability is at most $e^{-C\log^2 n}$; we will later keep this convention without a mention).
\begin{center}
\begin{figure*}[!ht]
$
\input xy
\xyoption{all}  \xymatrix@R=0.05in@C=1.3in{
\framebox{\mbox{\bf Prover}}((s_1, s_2), u|t)&\ar@{}[l] \framebox{\mbox{\bf Verifier}}(u|t) \\
\txt<12pc>{${\bf y}_{1}, {\bf y}_{2}\leftarrow {\cal Y}^\mu, {\bf v}= a{\bf y}_{1}+{\bf y}_{2}$} \ar[r]^-{{\bf v}} &*\txt<12pc>{} \\
\txt<12pc>{$\underline{b=1, 2}: \ \ {\bf z}_b=s_bc\cdot {\bf 1}+{\bf y}_b\quad\quad $\\
$A=\{j\mid z_{1j}, z_{2j}\in {\cal Z}\},\ j^*\leftarrow A$\\
$\underline{b=1, 2}:\  z_b=z_{bj^*}+\sum_{j\ne j^*} y_{bj}$ }& \txt<12pc>{$c\leftarrow {\cal C}$}\ar[l]_{c}  \\
\txt<12pc>{}\ar[r]^{z_1, z_2} &
\txt<12pc>{ $\underline{b=1, 2}: \quad ||z_b||_\infty<\eta_t?$\\
$\sum_{j=1}^\mu v_j\stackrel{?}{=}az_1+z_2-uc$}
}
$
 \caption{Our Lattice-based    ID Scheme ({\tiny     {\bf Note}: Membership checks $c\in {\cal C}$ at Prover is  important but omitted in the figure; ${\bf 1}$ is the vector of all 1 of length $\mu$.}) }
\label{fig: Ring-LWE}
\end{figure*}
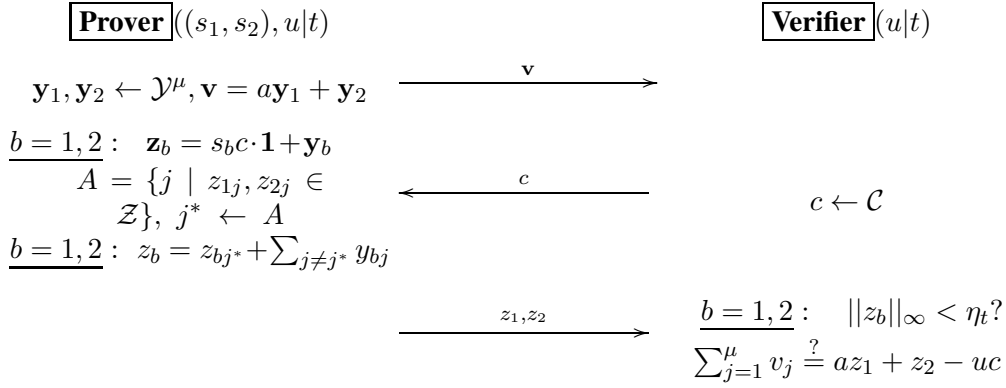
\end{center}
\subsubsection{Analysis}

In this section, we analyze our ID scheme. We start with some preparations. The following lemma is adapted from \cite[Lemma 4]{AF+12}, where our restriction that  the element $c$ of ${\cal C}$ has a degree at most $n/2$, does not affect the proof.

\vspace{.05in} \begin{lemma} \cite{AF+12} If $s\leftarrow D_{R, \sigma}$ and  $c\leftarrow {\cal C}$, then
\begin{align*}
\Pr(||sc||_\infty\le \sigma {n}^{1/2}\log^3n)\ge 1-e^{-\Omega(\log^2n)},
\end{align*}
where $e=2.71828\cdots$ is the Euler's number  \label{le: sc}
\end{lemma}

\vspace{.05in} The lemma below was in  the proof of \cite[Lemma 3]{AF+12}.
\begin{lemma} \cite{AF+12} Fix $\gamma \in R$ with $||\gamma||_\infty\le \sigma {n}^{1/2}\log^3n$. Then, for  $y\leftarrow {\cal Y}$, we have
\begin{align*}
&\Pr(\gamma+y\in {\cal Z})\ge \frac{1}{e}-\frac{1}{en}\\
& \Pr(\gamma +y=g\mid \gamma +y\in {\cal Z})=\frac{1}{|{\cal Z}|}, \forall g\in {\cal Z}.
\end{align*}
 \label{le: sc+y}
\end{lemma}

\begin{lemma} Let $A$ be the index set in our ID scheme. Then, $\Pr(A=\emptyset)<e^{-\Omega(\log^2 n)}$ for $\mu\ge \Omega(\log^2n)$.  \label{le: A}
\end{lemma}
\noindent {\bf Proof. } Notice $z_{bj}=sc+y_{bj}$ for $b=1, 2.$ By Lemma \ref{le: sc}, $||sc||_\infty\le \sigma{n}^{1/2}\log^3n$ with probability $1-e^{-\Omega(\log^2 n)}.$ Fixing $sc$ (that satisfies  this condition), $z_{bj}$ for $b=1, 2, j=1, \cdots, \mu$ are independent and thus by Lemma \ref{le: sc+y}, $A=\emptyset$ with probability at most $(1-\frac{1}{e^2}(1-\frac{1}{n})^2)^\mu<(1-\frac{1}{4e^2})^\mu, $ exponentially small. Together with the probability for $||sc||_\infty\le \sigma n^{1/2}\log^3n$,  we conclude the lemma. $\hfill\square$

\vspace{.05in} \begin{lemma}  If $u\leftarrow {\cal C}$, then  $u$  is invertible in $R_q$ with probability  $1-(1+2\log n)^{-n/2}.$ \label{le: c-c'}
\end{lemma}
\noindent {\bf Proof. } Recall that  $q\equiv 3$ mod $8$ in this section.  By Blake et al. \cite[Theorem 1]{BGM93}, $x^n+1=\Phi_1(x)\Phi_2(x)$ mod $q$, where $\Phi_1(x), \Phi_2(x)$ have  degree $n/2$ and are   irreducible over $\mathbb{Z}_q$. By Chinese remainder theorem, $u$ is invertible in $R_q$ if and only if it is non-zero mod  $\Phi_b(x)$ for both $b=1,2$. Since $u$ has a degree at most $n/2-1$, $u$ remains unchanged  after mod $\Phi_b(x)$.  Hence, it is invertible in $R_q$ if and only if $u$ is non-zero. This has a probability $1-(1+2\log n)^{-n/2}.$     $\hfill\square$

\vspace{.10in} \noindent {\bf Simulability. } We now show the simulability of our ID scheme.  Given the public-key $u$ and $c\leftarrow {\cal C}$, we define the simulator ${\bf SIM}$ as follows.
\begin{itemize}
\item[-] Sample $j^*\leftarrow [\mu]$ and  $z_{1j^*}, z_{2j^*}\leftarrow {\cal Z}$; compute $v_{j^*}=az_{1j^*}+z_{2j^*}-uc;$
\item[-] For $j\in [\mu]-\{j^*\}$, sample $y_{1j}, y_{2j}\leftarrow {\cal Y}$ and compute $v_{j}=ay_{1j}+y_{2j}.$
\item[-] Compute $z_b=z_{bj^*}+\sum_{j\ne j^*} y_{bj}, b=1, 2.$
\item[-] Output ${\bf v}=(v_1, \cdots, v_\mu)^T$ and $z_1, z_2.$
\end{itemize}
This simulation is valid by the following lemma.

\begin{lemma} The output of ${\bf SIM}$ is statistically close to the real transcript, even if the distinguisher has oracle access to ${\cal O}((s_1, s_2), u)$, where  $(s_1, s_2)\leftarrow D_{R, \sigma}^2$ is the private key and $u=as_1+s_2$ is the public-key.
\end{lemma}

\noindent{\bf Proof. } First,  we can assume $A\ne \emptyset$ for  the real transcript as by Lemma \ref{le: A} this is violated negligibly only. Then, by symmetry, $j^*$ for the real transcript is uniformly random over $\{1, \cdots, \mu\}.$ By the definition of $j^*$, we know that $z_{1j^*}, z_{2j^*}$ both belong to ${\cal Z}$. In this case, by Lemma \ref{le: sc+y}, $sc+y_{1j^*}, sc+y_{2j^*}$ for the real transcript with given $sc$ satisfying $||sc||_\infty<\sigma{n}^{1/2}\log^3 n$, are independent and uniformly random over ${\cal Z}$.
By lemma \ref{le: sc}, we conclude that $z_{1j^*}$ and $z_{2j^*}$ are statistically close to uniform over ${\cal Z}$ if they belong to ${\cal Z}$. On the other hand, when $z_{1j^*}$ and $z_{2j^*}$ are given, $v_{j^*}$ is fixed as $v_{j^*}=az_{1j^*}+z_{2j^*}-uc$. Thus, our simulation of $z_{1j^*}, z_{2j^*}, v_{j^*}$ is statistically close to that in the real transcript. On the other hand, our simulation of $y_{1j}, y_{2j}, v_j$ for $j\ne j^*$ is exactly according to the real distribution. Thus, our simulation is statistically close to the real transcript. This closeness holds (even given  adversary view, which includes the responses from ${\cal O}_{id}$). Hence, the simulability  follows.  $\hfill\square$

\vspace{.05in} \noindent {\bf Security. } Now we prove the security of our ID scheme, where the attacker needs to generate $z_1, z_2$ (given challenge $c$) to pass  the verification w.r.t. an aggregated public-key $\overline{u}$. We show that this is unlikely by  the ring-SIS assumption.

\vspace{.05in} \begin{lemma} Under ring-LWE$_{q, \sigma, 2n}$ and ring-SIS$_{3, q, \beta_{t^*}}$ assumptions, our scheme is $t^*$-secure (with respect to   Definition \ref{def: IDsec}), where $\beta_{t^*}=16\eta_{t^*}\sqrt{n}\log^2 n$ and $\sigma=\Omega(n)$.
\end{lemma}
\noindent {\bf Proof. } If there exists an adversary ${\cal D}$ that breaks our ring-based  ID scheme with non-negligible probability $\epsilon$, then we construct an adversary ${\cal A}$ that breaks ring-SIS assumption with a non-negligible probability $\epsilon'$. The idea is to make use of ${\cal D}$ to construct an algorithm $\textsf{A}$ for the nested forking lemma and then uses the output of the forking algorithm to obtain a solution for ring-SIS problem. Upon a challenge $u_1$ and $a$ (both uniformly over $R_q$), ${\cal A}$ needs to find short $\alpha_1, \alpha_2, \alpha_3\in R$ so that $a\alpha_1+\alpha_2+u_1\alpha_3=0.$ Toward this, ${\cal A}$ constructs an algorithm $\textsf{A}((u_1, a), \lambda_1, c; \rho)$ as follows (so $q=2$ in the forking algorithm), where $\lambda_i, c\leftarrow {\cal C}$ and $\overline{u}=\sum_{i=1}^t \lambda_i\cdot u_i$ with $u_i\in R_q$ (in the description of $\textsf{A}$).

\vspace{.1in} \noindent {\bf  Algorithm} $\textsf{A}((u_1, a),\lambda_1, c; \rho)$

{\bf Parse} $\rho$ as two parts: $\rho=\rho_0|\rho_1$

$(st_0, u_2, \cdots, u_{t})\leftarrow {\cal D}(u_1, a; \rho_0)$

$\lambda_2, \cdots, \lambda_t\leftarrow {\cal C}$ using randomness $\rho_1$

$st_1|{\bf v}\leftarrow {\cal D}(st_0, \lambda_1, \cdots, \lambda_t)$;

 $(z_1, z_2)\leftarrow {\cal D}(st_1, c)$;

{\bf If} $||z_b||_\infty<\eta_t$ {\bf and} $\sum_{j=1}^\mu v_j =az_1+z_2-\overline{u}c$,  {\bf then}

\quad \hspace{.24in}  $b=1$;

{\bf else} \hspace{.1in} $b=0$;

{\bf Output} $(b, 2b, \{u_i|\lambda_i\}_1^t|{\bf v}|z_1|z_2|c|a).$

\vspace{.1in}\noindent  \vspace{.05in} \noindent From the description of $\textsf{A}$ and the forking algorithm $F_A$ (for the forking lemma),  the rewinding in  $F_A$ only updates $\lambda_1$ and/or $c$ as well as variables affected by $(\lambda_1, c)$. In terms of forking lemma terminology, we have $(h_1, h_2)=(\lambda_1, c)$ and $I_0=1, J_0=2$ (for a successful execution; otherwise, $\textsf{A}$ will abort when $I_0\le J_0$).   Let us now analyze algorithm forking algorithm $F_{\textsf{A}}.$ When four executions are executed successfully (i.e., $b=1$ for all cases), then the output for each execution will be described as follows.
\begin{itemize}
\item[-] {\em Execution 0. } It outputs $(1, 2, \{u_i|\lambda_i\}_1^t|{\bf v}|z_1|z_2|c|a)$.  Since it succeeds, $||z_b||_\infty\le \eta_t\ (b=1, 2)$ and
\begin{equation}
\sum_{i=1}^\mu v_i =az_1+z_2-\overline{u}c.   \label{eq: DL-0}
\end{equation}

\item[-] {\em Execution 1. } Compared with execution 0, the input only changes $c$ to $\hat{c}$. From the code of $\textsf{A}$, the output is
$(1, 2, \{u_i|\lambda_i\}_1^t|{\bf v}|\hat{z}_1|\hat{z}_2|\hat{c}|a)$.
Since it  succeeds, $||\hat{z}_b||_\infty\le \eta_t\ (b=1, 2)$ and
\begin{equation}
\sum_{i=1}^\mu v_i =a\hat{z}_1+\hat{z}_2-\overline{u}\hat{c}. \label{eq: DL-1}
\end{equation}

\item[-] {\em Execution 2. } Compared with execution 0, the input changes $\lambda_1$ to $\bar{\lambda}_1$ and changes $c$ to $\bar{c}$. From the code of $\textsf{A}$, the output is
$(1, 2, \{u_i|\lambda_i\}_2^t|u_1|\bar{\lambda}_1|{\bf v}'|\bar{z}_1|\bar{z}_2|\bar{c}|a)$.
Since it succeeds, $||\bar{z}_b||_\infty\le \eta_t\ (b=1, 2)$ and
\begin{equation}
\sum_{i=1}^\mu v_i' =a\bar{z}_1+\bar{z}_2-\overline{u'}\bar{c},   \label{eq: DL-2}
\end{equation}
where $\overline{u'}=\bar{\lambda}_1u_1+\sum_{i=2}^t \lambda_i u_i.$

\item[-] {\em Execution 3. } Compared with execution 0, the input changes $\lambda_1$ to $\bar{\lambda}_1$ and changes $c$ to $\underline{c}$. From the code of $\textsf{A}$, the output is $(1, 2, \{u_i|\lambda_i\}_2^t|u_1|\bar{\lambda}_1|{\bf v}'|\underline{z}_1|\underline{z}_2|\underline{c}|a)$.
    Since it  succeeds, $||\underline{z}_b||_\infty\le \eta_t\ (b=1, 2)$ and
\begin{equation}
\sum_{i=1}^\mu v_i' =a\underline{z}_1+\underline{z}_2-\overline{u'}\underline{c}.  \label{eq: DL-3}
\end{equation}
\end{itemize}
From Eqs. (\ref{eq: DL-3})(\ref{eq: DL-2}), ${\cal A}$ can derive
\begin{equation}
a(\underline{z}_1-\bar{z}_1)+(\underline{z}_2-\bar{z}_2)-\overline{u'}(\underline{c}-\overline{c})=0.  \label{eq: DL-4}
\end{equation}
From Eqs. (\ref{eq: DL-1})(\ref{eq: DL-0}),
\begin{equation}
a(\hat{z}_1-z_1)+(\hat{z}_2-z_2)-\overline{u}(\hat{c}-c)=0. \label{eq: DL-5}
\end{equation}
Notice that Eq. (\ref{eq: DL-4})$\times (\hat{c}-c)$-Eq. (\ref{eq: DL-5})$\times (\underline{c}-\overline{c})$ gives
\begin{equation}
a\alpha_1+\alpha_2-{u}_1\alpha_3=0,
\end{equation}
where
\begin{align}
\alpha_1=&(\underline{z}_1-\bar{z}_1)(\hat{c}-c)-(\hat{z}_1-z_1)(\underline{c}-\overline{c})\\
\alpha_2=& (\underline{z}_2-\bar{z}_2)(\hat{c}-c)-(\hat{z}_2-z_2)(\underline{c}-\overline{c})\\
\alpha_3=& (\lambda_1-\bar{\lambda}_1)(\hat{c}-c)(\underline{c}-\overline{c}).
\end{align}
Hence, $(\alpha_1, \alpha_2, -\alpha_3)$ forms a solution to ring-SIS problem with parameter $(a, 1, u).$ It suffices to verify that each $\alpha_i$ is short and also at least one of them is non-zero. For the second condition, it suffices to make sure that the probability for $\alpha_3=0$ is small. Notice that by Chinese remainder theorem,  $\alpha_3=0$ implies $\lambda_1=\bar{\lambda}_1$ mod $\Phi_1(x)$ or $\underline{c}=\overline{c}$ mod $\Phi_1(x)$ or $\hat{c}=c$ mod $\Phi_1(x).$ Similarly, this must also hold for modular $\Phi_2(x)$ but it suffices to consider $\Phi_1(x)$ only. Since $\lambda_1, \bar{\lambda}_1, \underline{c}, \overline{c}, \hat{c}$ is uniformly random over ${\cal C}$, each of the equality holds with probability $(1+2\log n)^{-n/2}$ only and hence $\Pr(\alpha_3=0)\le 3(1+2\log n)^{-n/2},$ negligible! For the first condition, notice that $||\hat{c}-c||_\infty\le 2\log n$ and $||\underline{z}_1-\overline{z}_1||_\infty\le 2\eta_t$. Further, the
constant term of $(\hat{c}-c)(\underline{z}_1-\overline{z}_1)$ is
\begin{equation*}
(\hat{c}-c)[0]\cdot (\underline{z}_1-\overline{z}_1)[0]-\sum_{k=1}^{\frac{n}{2}-1} (\hat{c}-c)[k]\cdot (\underline{z}_1-\overline{z}_1)[n-k]
\end{equation*}
which, by Heoffding inequality, has an absolute value at most $\sqrt{n/2}\log n\cdot 8\eta_t\log n\le  8\eta_t\sqrt{n}\log^2 n$, with probability at least $1-e^{-\Omega(\log^2n)}.$ The constant term of $(\hat{z}_1-z_1)(\underline{c}-\overline{c})$ is similar. Hence, $|\alpha_1[0]|\le 16\eta_t\sqrt{n}\log^2 n$, with probability at least $1-e^{-\Omega(\log^2n)}.$   The general case of  $\alpha_1[i]$ is similar.
Hence, $||\alpha_1||_\infty\le 16\eta_t\sqrt{n}\log^2 n$ with probability $1-e^{-\Omega(\log^2n)}.$ Similarly, $||\alpha_2||_\infty$ has the same property. We can use the above proof  technique to show that $||(\hat{c}-c)(\underline{c}-\overline{c})||_\infty\le 8\log n\cdot \sqrt{n}\log^2 n$ with probability $1-e^{-\Omega(\log^2n)}.$ Since $\lambda_1, \bar{\lambda}_1$ is uniformly random over ${\cal C}$, using the same technique, we have $||\alpha_3||_\infty\le \sqrt{n}\log n\cdot 32\sqrt{n}\log^4 n=32n\log^5 n,$ with probability $1-e^{-\Omega(\log^2 n)}.$ Thus, we find a ring-SIS solution $(\alpha_1, \alpha_2, -\alpha_3)$ of length at most $16\eta_t\sqrt{n}\log^2 n$. Assume that the probability that ${\cal D}$ succeeds in one execution is $\hat{\epsilon}$. Then, by forking lemma, it succeeds in four executions with probability $\hat{\epsilon}^4-3(1+2\log n)^{-n/2}.$ This implies that ${\cal A}$ breaks the ring-SIS assumption with probability at least $\hat{\epsilon}^4-3(1+2\log n)^{-n/2}-e^{-\Omega(\log^2 n)}.$

Finally, notice that the input $u_1$ is uniformly random over $R_q$ while in our ID scheme $u_1=as_1+s_2$ for $s_1, s_2\leftarrow D_{R, \sigma}. $  However, under ring-LWE assumption, it is immediate that $\hat{\epsilon}\ge \epsilon-\textsf{negl}(n).$ Hence, ${\cal A}$ can succeed with probability at least ${\epsilon}^4-\textsf{negl}(n),$  this contradicts the assumption of ring-SIS. $\hfill\square$

\vspace{.05in} \noindent {\bf Linearity. } Let ${\cal SK}={\cal RSP}=(R_q, R_q), {\cal CMT}=R_q^\mu, {\cal PK}=R_q, {\cal R}=R_q$. We now verifies the linearity.
\begin{itemize}
\item[i.] Obviously, ${\cal SK}$ is a ${\cal R}$-module under the operation $\bullet$: for $(s_1, s_2)\in {\cal SK}$ and $c\in {\cal R}$, $c\bullet(s_1, s_2)=(cs_1, cs_2)$, where $cs_1$ and $cs_2$ are multiplications in $R_q.$ Other cases are similar.

\item[ii.]  If $(s_{1i}, s_{2i})\in {\cal SK}$ and $\lambda_i \in {\cal C}$ for $i=1, \cdots, t$, then $\sum_{i=1}^t (\lambda_is_{1i}, \lambda_is_{2i})=(\sum_{i=1}^t \lambda_is_{1i}, \sum_{i=1}^t\lambda_is_{2i})$ is obviously the private key of
$\sum_{i=1}^t \lambda_i \cdot (as_{1i}+s_{2i})=a(\sum_{i=1}^t \lambda_i s_{1i})+(\sum_{i=1}^t \lambda_i s_{2i}).$ However, we emphasize that this key  is not necessarily short.  But for  randomly  generated $(pk_i, sk_i, \lambda_i)$'s, Lemma \ref{le: Z} implicitly implies that the aggregated private key has length at most $ 2\sqrt{nt}\sigma\log^3 n$ (except for probability $e^{-\Omega(\log^2 n)}$); see $\max_v |S_v|$ with $|S_v|$ given in the proof of Lemma \ref{le: Z}).

\item[iii.] If $\{({\bf v}_i, c, z_{1i}, z_{2i})\}_{i=1}^t$ are {\em honestly} generated accepting transcripts w.r.t the honestly public/private key pairs $\{(u_i, (s_{1i}, s_{2i}))\}_i$, then  \remove{by correctness, except for probability $e^{-\Omega(\log^2 n)}$, we have}
\begin{align}
&\sum_{j=1}^\mu v_{ij}=az_{1i}+z_{2i}-u_ic. \label{eq: 40}
\end{align}
Together with Lemma \ref{le: Z} below, for $h_1, \cdots, h_t\leftarrow {\cal C}$,  $(\sum_{i=1}^t h_i{\bf v}_i, c, \sum_{i=1}^t h_i z_{1i}, \sum_{i=1}^t h_i z_{2i})$ satisfies (except for probability $e^{-\Omega(\log^2 n)}$)
\end{itemize}
\begin{align*}
\ \ \ ||\sum_{i=1}^th_i z_{1i}||_\infty\le& \eta_t,\quad ||\sum_{i=1}^t h_i z_{1i}||_\infty\le \eta_t,\\
\ \ \  \sum_{j=1}^\mu (\sum_{i=1}^t h_iv_{ij})=&a(\sum_{i=1}^t h_i z_{1i})+(\sum_{i=1}^t h_i z_{2i})-(\sum_{i=1}^t h_iu_i)c,
\end{align*}
\ \ \ \ where $\eta_t=5\sigma n^2\sqrt{t\mu}\log^6 n.$
The linearity follows.

\begin{lemma} Fix integer $t\ge 2$ and $\sigma\ge \omega(\log {n}).$  Assume $s_i\leftarrow D_{R, \sigma}, h_i\leftarrow {\cal C}, y_{ij}\leftarrow {\cal Y}$ for $i\in [t], j\in [\mu], c\leftarrow {\cal C}$. Let
\begin{equation}
Z=\sum_{i=1}^t h_i(s_ic+\sum_{j=1}^{\mu} y_{ij}).
\end{equation}
 Then, $||Z||_\infty\le \eta_t$ with probability $1-e^{-\Omega(\log^2 n)}.$ \label{le: Z}
\end{lemma}
\noindent {\bf Proof. } Notice
\begin{equation*}
Z[0]=\sum_{v=0}^{n-1} S_v\cdot c[v]-\sum_{i=1}^t\sum_{k=0}^{n-1} h_i[n-k]\cdot Y_{ik},
\end{equation*}
 where
$Y_{ik}=\sum_{j=1}^\mu y_{ij}[k]$, $h_i[n]\stackrel{def}{=}-h_i[0]$ and
$$S_v=\sum_{i=1}^t\sum_{k=0}^{n-1} h_i[n-k]s_i[k-v].$$ By \cite[Lemma 4.2]{GPV08}, $||s_i||_\infty \le \sigma\log n$, except for probability $e^{-\Omega(\log^2n)}$. When this is satisfied, terms $h_i[n-k]s_i[k-v]$ in $S_v$ are independent random variables in the range $[-\sigma\log^2 n, \sigma\log^2 n].$ By Heoffding inequality, $|S_v|\le 2\sqrt{nt} \sigma\log^3 n$, except for probability $e^{-\Omega(\log^2n)}.$ Since $y_{ij}[k]$ is uniformly random over $[-\sigma n^{1.5}\log^3 n, \sigma n^{1.5}\log^3 n]$, by  Heoffding inequality, $|Y_{ik}|\le 2\sigma \sqrt{\mu} n^{1.5}\log^4 n$, except for a probability $e^{-\log^2 n}.$ Assuming these inequalities for $S_v$ and $Y_{ik}$, we know that from Heoffding inequality again,
\begin{align*}
&|\sum_{v=1}^{n-1} S_v\cdot c[v]|\le 4\sigma n\sqrt{t}\log^5 n\\
&|\sum_{i, k} h_i[n-k]\cdot Y_{ik}|\le \sqrt{n t}\log n\cdot 4\sigma \sqrt{\mu} n^{1.5}\log^5 n,
\end{align*}
except for probability $e^{-\Omega(\log^2 n)}.$  Hence, we conclude that
$|Z[0]|\le 5\sigma n^2\sqrt{t\mu}\log^6 n$, except for $e^{-\Omega(\log^2 n)}.$ We can similarly bound $Z[i]$ for $i\ge 1$ and so
 $||Z||_\infty \le 5\sigma n^2\sqrt{t\mu}\log^6 n$, except for probability $e^{-\Omega(\log^2 n)}.$ $\hfill\square$

\vspace{0.05in}  \subsubsection{Key-and-Signature Compact Multi-signature Scheme from our ID scheme.} \quad With the simulability, linearity and security for our ID, we can use our compiler to convert it  into a secure multi-signature. We now describe this scheme as follows.

Let $(s_{1i}, s_{2i})$ be the private key of public-key $u_i=as_{1i}+s_{2i}$ for $i=1, \cdots, t$. If the users of $u_1, \cdots, u_t$ want to jointly sign $M$, they compute the aggregated public-key $\overline{u}|t$ and execute the protocol  as follows, where $H_0, H_1: \{0, 1\}^*\rightarrow {\cal C}$ and we define $\overline{w}=\sum_{i=1}^t H_0(u_i, U) w_i$ for any  list of variables   $w_1, \cdots, w_t$ in the description and $U=(u_1, \cdots, u_t)$ (e.g., $\overline{\bf v}=\sum_{i=1}^t H_0(u_i, U)\cdot  {\bf v}_i$).

\begin{itemize}
\item[\textbullet] {\bf R-1}. \quad User generates ${\bf y}_{1i}, {\bf y}_{2i}\leftarrow {\cal Y}^\mu$, computes  ${\bf v}_i=a{\bf y}_{1i}+{\bf y}_{2i}$ and sends $H_0({\bf v}_i|u_i)$ to other users.

\item[\textbullet] {\bf R-2}. \quad Upon receiving all $r_j, j=1, \cdots, t$, user $i$ sends ${\bf v}_i$ to other users.

\item[\textbullet] {\bf R-3}.  \quad Upon all ${\bf v}_j$, user $i$ verifies its consistency with $r_j$. If verification fails, it rejects; otherwise, it computes $\overline{\bf v}$ and $c=H_1(\overline{u}|\overline{\bf v}|M)$ as well as the  response $(z_{1i}, z_{2i})$ for  challenge $c$ in the ID scheme with committing message ${\bf v}_i$.
\item[\textbullet] {\em output}. \quad After receiving $(z_{1j}, z_{2j})$ for $j\in [t]$, user $i$ computes multi-signature $(\overline{z}_1, \overline{z}_2,  \overline{\bf v})$. The aggregated public-key is $\overline{u}|t.$
\item[\textbullet] {\em Verify}. \quad Upon $(\overline{z}_1, \overline{z}_2, \overline{\bf v})$,  it verifies the following with $\overline{u}|t$ and accepts only if it is valid:
\end{itemize}
\begin{align}
&||\overline{z}_{1}||_\infty\le \eta_t,\quad ||\overline{z}_{2}||_\infty\le \eta_t,\\
&\sum_{j=1}^\mu \overline{v}_{j}=a\overline{z}_{1}+\overline{z}_{2}-\overline{u}c,
\end{align}
where $\eta_t=5\sigma n^2\sqrt{t\mu}\log^6 n.$
Denote this multi-signature scheme by \textsf{RLWE-MultiSig}.  From our compiler and the properties of our ID scheme, we obtain the following.

\begin{corollary} Let $\eta_{t^*}=5\sigma n^2\sqrt{t^*\mu}\log^6 n$, $\sigma=\Omega(n)$ and $\beta_{t^*}=16\eta_{t^*}\sqrt{n}\log^2 n$. Under Ring-LWE$_{q, \sigma, 2n}$ and Ring-SIS$_{3, q, \beta_{t^*}}$ assumptions, \textsf{RLWE-MultiSig} is $t^*$-EU-CMA secure. Especially, if the assumptions hold for $t^*=2^{\sqrt[4]{n}}$, then \textsf{RLWE-MultiSig} is EU-CMA secure.
\end{corollary}

\noindent {\bf Remark. }  As the best algorithm \cite{LM06,CDW17} can only solve ring-SIS$_{q, 3, \beta}$ with $\beta=2^{\tilde{O}(\sqrt{n})}$, it is safe to assume ring-SIS$_{q, 3, \beta}$ with any polynomial $\beta$. If the assumption is sound for $\beta=2^{\sqrt[4]{n}}$, our multi-signature scheme is EU-CMA secure, as $t^*$ can only be polynomial for a PPT adversary.

\section{Conclusion}
In this paper, we proposed a compiler that converts a type of identification scheme to a key-and-signature compact multi-signature. This special type of  ID owns a linear property. The aggregated public-key and multi-signature are of size both independent of the number of signers. We formulated this compiler through linear ID via the language of  ${\cal R}$-module and proved the security through a new forking lemma called nested forking lemma.
Under our compiler, the compact multi-signature problem has been reduced from a multi-party problem to a two-party problem. We realized our compiler with Schnorr ID scheme and a new lattice-based scheme. Our lattice multi-signature is the first of its kind that is key-and-signature compact without a restart  in the signing process.


\ifCLASSOPTIONcaptionsoff
  \newpage
\fi



%

\end{document}